\documentclass[fleqn,usenatbib]{mnras}

\usepackage{newtxtext} 
\usepackage[scaled=0.92]{helvet}

\usepackage[T1]{fontenc}
\usepackage{ae,aecompl}

\usepackage{graphicx,epstopdf}	
\usepackage{amsmath}	
\usepackage{amssymb}	

\newcommand{\Msun}{$\mathrm{M}_\odot$}
\newcommand{\Rsun}{$\mathrm{R}_\odot$}

\newcommand{\appropto}{\mathrel{\vcenter{
  \offinterlineskip\halign{\hfil$##$\cr
    \propto\cr\noalign{\kern2pt}\sim\cr\noalign{\kern-2pt}}}}}

\title[ECSNe and Low-mass CCSNe]
{Synthetic observables for electron-capture
supernovae and low-mass core collapse supernovae}

\author[A. Kozyreva et al.]{Alexandra~Kozyreva$^{\,1,2}$\thanks{E-mail: sasha@mpa-garching.mpg.de},
Petr Baklanov$^{\,3,4}$, Samuel Jones$^{\,5}$
\newauthor
Georg Stockinger$^{\,1,6}$,
Hans-Thomas Janka$^{\,1}$\\
\\
$^{1}$Max-Planck-Institut f\"ur Astrophysik, Karl-Schwarzschild-Str. 1, 85748 Garching, Germany,\\
$^{2}$Alexander von Humboldt Fellowship\\
$^{3}$NRC ``Kurchatov Institute'' -- ITEP, Moscow, 117218, Russia \\
$^{4}$National Research Nuclear University Moscow Engineering Physics Institute, Moscow, 115409, Russia\\
$^{5}$X Computational Physics (XCP) Division and Center for Theoretical Astrophysics (CTA), Los Alamos National Laboratory, Los Alamos, NM 87545, USA\\
$^{6}$Physik-Department, Technische Universit\"{a}t M\"{u}nchen, James-Franck-Str. 1, 85748 Garching,
Germany\\
}
\date{Accepted XXX. Received YYY; in original form ZZZ}

\pubyear{2020}

\begin{document}
\label{firstpage}
\pagerange{\pageref{firstpage}--\pageref{lastpage}}
\maketitle


\begin{abstract}
Stars in the mass range from 8~\Msun{} to 10~\Msun{} are expected to produce
one of two types of supernovae (SNe), either electron-capture supernovae (ECSNe) 
or core-collapse supernovae (CCSNe), depending on their previous evolution. 
Either of the associated progenitors retain extended and massive hydrogen-rich envelopes, 
the observables of these SNe are, therefore, expected to be similar.
In this study we explore the differences in these two types of SNe.
Specifically, we investigate three different progenitor models: a solar-metallicity ECSN
progenitor with an initial mass of 8.8~\Msun{}, a zero-metallicity progenitor with
9.6~\Msun{}, and a solar-metallicity progenitor with 9~\Msun{}, 
carrying out radiative transfer simulations for these progenitors.
We present the resulting light curves for these models. 
The models exhibit very low photospheric velocity variations of about 2000\,km\,s$^{\,-1}$, therefore, this may serve
as a convenient indicator of low-mass SNe. The ECSN has very unique light curves in broad bands,
especially the \emph{U} band, and does not resemble any currently observed SN.
This ECSN progenitor being part of a binary will lose its envelope for which
reason the light curve becomes short and undetectable.
The SN from the 9.6~\Msun{} progenitor exhibits also quite an unusual 
light curve, explained by the absence of
metals in the initial composition. The artificially iron polluted 
9.6~\Msun{} model demonstrates light curves closer to normal SNe~IIP.
The SN from the 9~\Msun{} progenitor remains the best candidate for so-called
low-luminosity SNe~IIP like SN~1999br and SN~2005cs.

\end{abstract}





\section[Motivation]{Introdution}
\label{sect:intro}

According to the Salpeter initial mass function, 
stellar populations are dominated by low-mass stars, with initial masses
below 8~\Msun{} \citep{1955ApJ...121..161S,2001MNRAS.322..231K}.
The majority of them live quiet, 
billion-year long lives if single and isolated, like our Sun. 
Some of them with initial masses 4\,--\,8~\Msun{} form degenerate carbon-oxygen cores, 
end up as white dwarfs and, if part of a binary, may become progenitors of 
thermonuclear supernovae \cite[SNe, see e.g.
][]{1980PASJ...32..303M,2000ARA&A..38..191H,2016A&A...589A..43N}. 
Massive stars with initial masses of 10\,--\,40~\Msun{} (referred to as
``typical'' massive) end their lives
in more spectacular explosions triggered by neutronisation of the iron-core
and subsequent gravitational collapse
\citep{2002RvMP...74.1015W,2003ApJ...591..288H,2016ApJ...818..124E,2016ApJ...821...38S}.

The narrow range of initial masses between 8~\Msun{} and 10~\Msun{} 
is thoroughly explored 
\citep[see e.g., ][]{2013ApJ...772..150J,2014ApJ...797...83J,2015MNRAS.446.2599D,2019ApJ...882..170J,2019MNRAS.484.3307M,2019ApJ...871..153T},
however, there is no clear understanding of the final remnants of these stars.
Evolution of these stars is determined by a number of shell-burning flashes, i.e.
it involves a complex history proceeding on both dynamical and nuclear 
time scales. Nevertheless, mass limits for this class of events are still a matter
of debate in the literature on stellar evolution modelling
\citep{2015MNRAS.447.3115J,2016A&A...586A.119S,2016MNRAS.456.1320T}.
Electron-capture SNe (ECSNe) occur for progenitors in the very narrow initial mass range of
about $\pm$0.2~\Msun{} around 8~\Msun{}, 7.8\,--\,8.2~\Msun{}, assuming single stellar evolution 
\citep{2007A&A...476..893S,2015MNRAS.446.2599D}. 
Progenitors in a wider initial stellar mass
range between 13.5~\Msun{} and 17.6~\Msun{} result in ECSNe while considering stellar evolution within
a binary \citep{2017ApJ...850..197P,2018A&A...614A..99S}. 
Consequently, this leads to a higher rate of ECSNe among core-collapse
explosions, although they might look as SNe~IIb or SNe~Ib/c \citep{2015MNRAS.451.2123T}.

Predicted observables depend strongly on the outcomes of stellar
evolution simulations, e.g. whether the envelope is hydrogen rich \citep{2009ApJ...705L.138P}, and the
post-explosion calculations, i.e. the
degree of macroscopic mixing. Therefore, the
real explosion can be adequately modelled as a thermal bomb.
However, details of the explosion simulations may strongly affect the distribution of 
chemical elements, erasing the chemically-stratified structure. For example,
strong mixing between chemically confined layers happens
when the reverse shock passes through the expanding stellar ejecta
accelerated by the forward shock. This may be caused by asymmetries which
occur at the earlier phase of the explosion, i.e. during shock revival.

It is worth mentioning that super-asymptotic-giant-branch (super-AGB) stars and those at the low-mass end of
the core-collapse SN (CCSN) progenitors are important contributors to
the chemical evolution of the Galaxy
\citep{2010MNRAS.403.1413K,2014MNRAS.437..195D,2016MmSAI..87..229K,2019ApJ...882..170J,2019IAUS..343..247K}. 
Namely, the main contribution of massive and super-AGB stars to Galactic
chemical evolution prior to
explosion is probably only isotopes produced in hot-bottom-burning and
some small subset of s-process isotopes.  If these stars are stripped being
part of a binary then they probably would not make either of these contributions.
If they explode as a core-collapse or a thermonuclear explosion,
they are major contributors of neutron-rich isotopes as {}$^{48}$Ca,
{}$^{50}$Ti, and {}$^{54}$Cr, {}$^{58}$Fe, {}$^{64}$Ni, {}$^{82}$Se, and
{}$^{86}$Kr and several isotopes beyond the iron-peak, e.g., Zn\,--\,Zr
\citep{2019ApJ...882..170J}.

In this study, fully self-consistent calculations
of three progenitors with initial masses 8.8~\Msun{}, 9~\Msun{}, and
9.6~\Msun{} are presented, including stellar evolution simulations, core-collapse
explosion simulations, and radiative transfer simulations. 
We consider our treatment as ``self-consistent'' in the sense that
the pre-collapse models from stellar evolution calculations were mapped into
PROMETHEUS-HOTB or PROMETHEUS-VERTEX to perform 3D explosion calculations
with a detailed treatment of the neutrino
energy deposition that triggers and powers the explosions, and then mapped 
1D-averages of the 3D explosion models into the radiation transport 
code STELLA without adding any additional explosion energy and {}$^{56}$Ni.
We present physically consistent calculations of:
\begin{enumerate}
\item stellar evolution from the zero-age main sequence (ZAMS) through the
nuclear burning stages until formation of the iron-core;
\item core-contraction, bounce, shock revival, finally SN explosion
until the shock breakout \citep{2020MNRAS.496.2039S};
\item hydrodynamics of the SN ejecta and evolution of the radiation field;
\item multi-band light curves and spectral energy distribution.
\end{enumerate}

The goal of the study is to explore differences in observables
for the progenitors in the narrow initial mass range of 8 to 10~\Msun{} which
may either be ECSNe or iron CCSNe.
We describe our models in Section~\ref{sect:method}, as well as our
methodology of computing
post-explosion hydrodynamical evolution and radiative transfer.
Section~\ref{sect:results} presents bolometric and broad band light curves
and their dependences on the progenitor metallicity, explosion energy,
hydrogen-to-helium ratio in the envelope, and the radius of the progenitor. 
We compare our simulations with
observed SNe in Section~\ref{sect:observe} 
trying to find any observed candidates matching our models. We present
our conclusions in Section~\ref{sect:conclusions}.


\section[Input models]{Input models and Method}
\label{sect:method}

\begin{figure}
\centering
\includegraphics[width=0.5\textwidth]{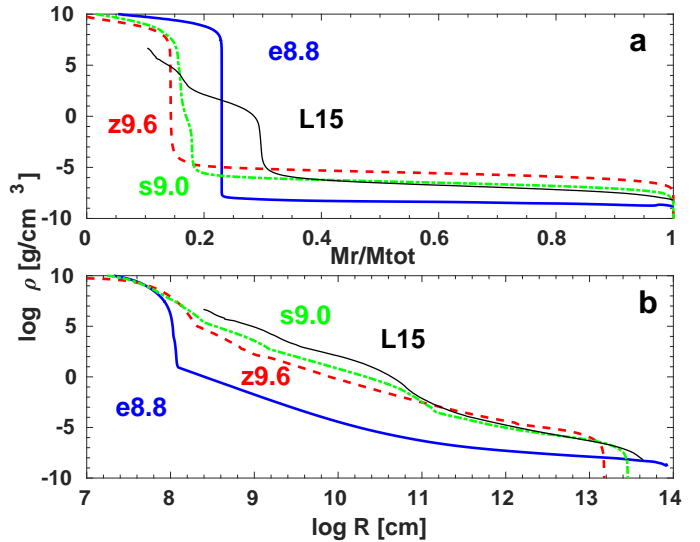}
\caption{Density structure of the models e8.8, z9.6, s9.0, and L15
prior to the explosion. The top plot shows the
density profiles versus mass coordinate relative to total mass. The
bottom plot presents the same density profiles versus radius.} 
\label{figure:density}
\end{figure}

\begin{table*}
\caption{Input models for our radiative transfer simulations. Note, that
both the e8.8 and s9.0 stellar models are initially at solar metallicity, but the
model e8.8 was computed with abundances from
\citet[][]{2009ARA&A..47..481A}, having metallicity Z$_\odot${}, and
the model s9.0, taken from \citet[][]{2016ApJ...821...38S}, was made with the
assumption of solar abundances from \citet{2003ApJ...591.1220L} having
metallicity Z$^{(1)}_\odot${}. ``Fe''
stands for the mass fraction of stable iron in the stellar envelope. ``E$_\mathrm{kin}$''
represents terminal kinetic energy of the entire ejecta.
``Time'' means time since bounce, i.e. the point where the models were mapped into
STELLA, and approximately corresponds to time of the shock breakout.
The comparison model L15 from \citet{2017ApJ...846...37U} is listed at the
bottom as well.}
\label{table:values}
\begin{tabular}{l|l|l|l|l|l|l|l|l}
\hline
  & M$_\mathrm{fin}$/M$_\mathrm{ej}$ [\Msun{}]& Z & X(Fe) & R [\Rsun{}] & M$_\mathrm{cut}$ [\Msun{}]
& Time [days] & E$_\mathrm{kin}^\mathrm{fin}$ [$10^{\,50}$~erg] & $^{56}$Ni [\Msun{}]\\
\hline
e8.8 3D & 5.83/4.5 & Z$_\odot${} & 4.7$\times 10^{\,-4}$ & 1200 & 1.326 & 5.2 & 0.86 & 0.0013 \\
\hline
Tracer  &      &   &   &&&&& 0.0077\\ 
        &      &   &   &&&&& 0.0141\\ 
\hline
Z-study &      & 0           & 0                     &      &       &     &      &\\
        &      & Z$_\mathrm{SMC}$& 1.4$\times 10^{\,-4}$ &      &       &     &      &\\
\hline
Energy-study & & & & & & & & \\
3e49-2D &      &   &                       &      &       &     & 0.3  & 0.0009 \\
6e49-2D &      &   &                       &      &       &     & 0.55 & 0.0011 \\
1e50-2D &      &   &                       &      &       &     & 0.92 & 0.0013 \\
1.5e50-2D &    &   &                       &      &       &     & 1.37 & 0.0012 \\
\hline
Radius-study & & & & & & & & \\
e8.8 evol & 5.82/4.5 & Z$_\odot${} & & 1200 & 1.326 &  & 0.86  & 0.00092 \\
e8.8 evol & 4/2.7 &  & & 900 &  &  & 0.86  & 0.00092 \\
e8.8 evol & 2.4/1.1 &  & & 600 &  &  & 0.86  & 0.00098 \\
e8.8 evol & 1.8/0.5 &  & & 400 &  &  & 0.86  & 0.00127 \\
\hline
z9.6    & 9.6/8.25  & 0 & 0                      & 214 & 1.353 & 5.8 & 0.81 & 0.0007 \\
        &      & Z$_\mathrm{SMC}$& 1.4$\times 10^{\,-4}$   &&&&&\\
        &      & Z$_\odot${}&5$\times 10^{\,-4}$     &&&&&\\
        &      & Z$^{(1)}_\odot${}&1.4$\times 10^{\,-3}$  &&&&&\\
\hline
s9.0    & 8.75/7.4 & Z$^{(1)}_\odot${} & 1.46$\times 10^{\,-3}$ & 409 & 1.356 & 4.2 & 0.68 & 0.0051 \\
\hline
L15-nu & 15/13    & Z$^{(1)}_\odot${} & 1.36$\times 10^{\,-3}$ & 627 & 2    & 0 & 5.5 & 0.036\\
L15-tb & 15/13.47 & Z$^{(1)}_\odot${} & 1.36$\times 10^{\,-3}$ & 627 & 1.53 & 0 & 5.4 & 0.036\\
\hline
\end{tabular}
\end{table*}

We use three self-consistently modeled SN simulations 
as presented by \citet{2020MNRAS.496.2039S} for this study, namely, 
the ECSN model e8.8, and two low-mass CCSN models z9.6 and s9.0. e8.8 is a
1D solar-metallicity stellar
evolution model which was constructed the following way before being mapped
into the \verb|PROMETHEUS| code: a 2.2~\Msun{} core was initially calculated by
\citet{1987ApJ...322..206N}. The envelope was computed with
\verb|MESA|\footnote{Modules for Experiments in Stellar Astrophysics
\url{http://mesa.sourceforge.net/}
\citep{2011ApJS..192....3P,2013ApJS..208....4P,2015ApJS..220...15P,2018ApJS..234...34P,2019ApJS..243...10P}.} 
\citep{2013ApJ...772..150J}, then truncated and attached to
the core that was slightly reduced in mass \citep[][]{2018SSRv..214...67N,2020ApJ...889...34L}.
The reduced envelope mass\footnote{8.8~\Msun{} model has the final mass of
8.544~\Msun{} \citep[see Table~1 in ][]{2013ApJ...772..150J}.} is 
explained by the pre-collapse stellar
evolution which contains numerous shell-burning flashes, i.e.,
pulsation-driven mass-loss episodes, and steady-state mass-loss
\citep{2008ApJ...675..614P}.
Furthermore, the available prescriptions for the mixing processes and
mass-loss remain the
overarching uncertainty in the final outcome of the models in the range between
8~\Msun{} and 10~\Msun{} \citep{2013ApJ...772..150J}. The final total mass of 5.83~\Msun{} was
chosen to match the estimated mass of the Crab Nebula 
\citep{1982A&A...110L...3H,1982Natur.299..803N,2013ApJ...771L..12T}. The
model e8.8 is relatively physically large, having a radius of 1200~\Rsun{} which is
similar to the average red supergiant. However, we show that the extended
tenuous envelope makes the observational properties of this particular model
quite unique.

\begin{figure*}
\centering
\large\bf\textsf{\input{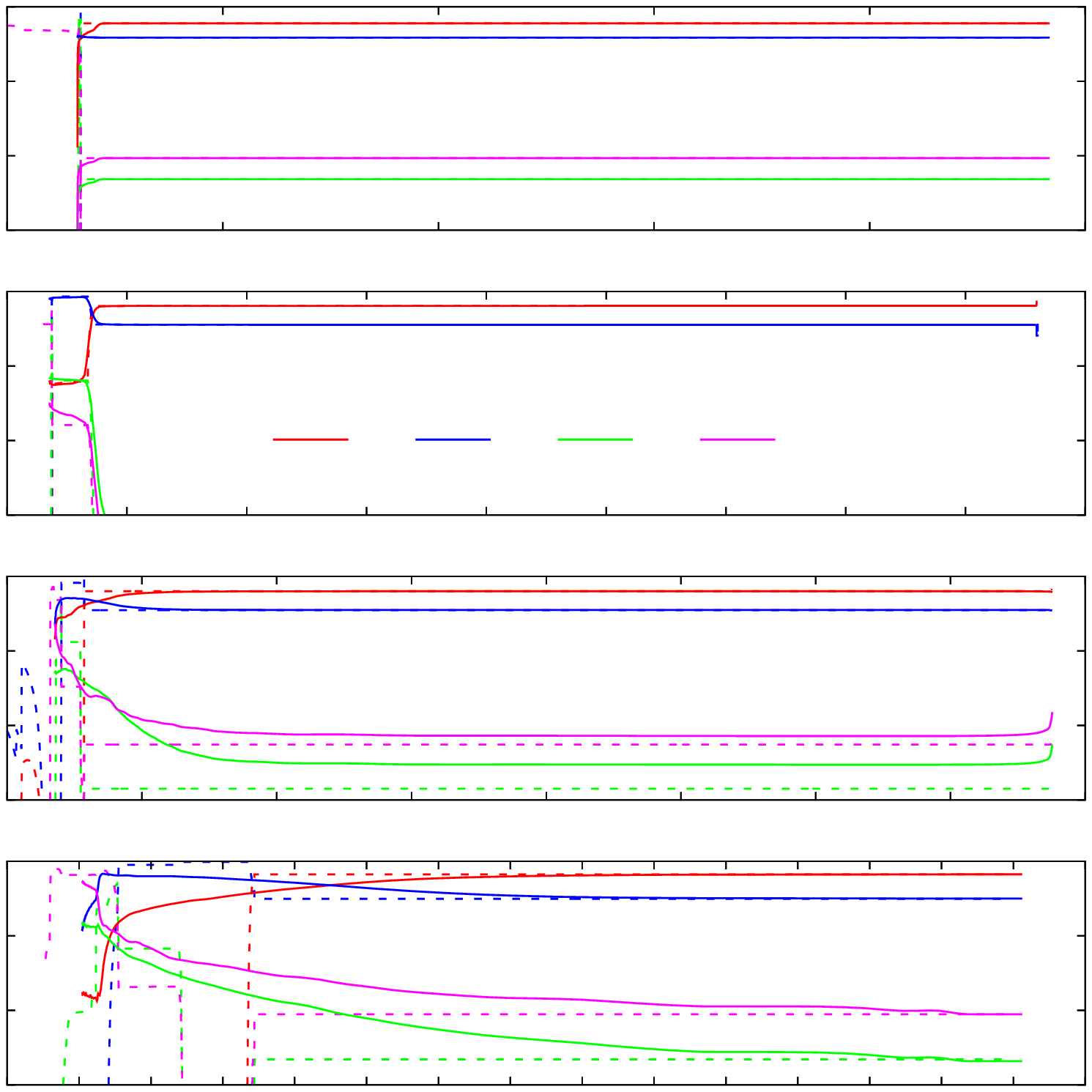}}
\caption{Chemical structure (hydrogen, helium, carbon, and oxygen profiles) of the
ejecta. Model designation as labeled. Dashed lines indicate the pre-collapse
chemical structures of the models while solid lines present post-explosion
structures.}
\label{figure:chemie}
\end{figure*}

Model z9.6 is a zero-metallicity stellar evolution model with initial mass of
9.6~\Msun{} calculated with the
\verb|KEPLER| code \citep{1978ApJ...225.1021W}. With no metals in the original chemical mixture this model
is relatively compact, its radius at the moment of iron-core
collapse being 214~\Rsun{}, i.e. the star is somewhere at the boundary between
blue and red supergiants \citep{2010ApJ...724..341H}. The solar-metallicity model s9.0 is
a star with initial 9~\Msun{} also produced with \verb|KEPLER|
\citep{2016ApJ...821...38S}.

The choice of the models in the present study is explained by the
following aspects:
\begin{enumerate}
\item One of the purposes of our study is to explore
differences and similarities in the resulting supernova observables between
models in the narrow range of initial ZAMS masses between 8~\Msun{} and 10~\Msun{}.
\item Pre-collapse model e8.8 is the latest version of a super-AGB ECSN progenitor
available to us.
\item The z9.6 progenitor has a density structure prior to the collapse similar to e8.8,
namely, a steep density gradient at the core-envelope interface (see
Figure~\ref{figure:density}). 
\item Pre-collapse model s9.0 is the representative of low-mass progenitors and explodes fairly easily in self-consistent
3D simulations, having similar energy as the other two cases 
\citep[see
][]{2017ApJ...850...43R,2019ApJ...873...45G,2020MNRAS.496.2039S}.
\end{enumerate}

We note that adding 3D simulations of more progenitors or more model variations is 
computationally expensive and currently not feasible. One of the reasons is the
long time, up to 5 days, required for the shock to break out from these very 
extended progenitors. The approximate time for the shock breakout is
definited as:
\begin{equation}
t_\mathrm{SBO}\sim 0.91~\mathrm{days}\,R_{500}\,E_{51}\,M_{10}\,,
\label{equation:tSBO}
\end{equation}
where 
$R_{500}=R / 500$~\Rsun{}, $E_{51}=E_\mathrm{exp} / 10^{\,51}$~erg, and
$M_{10}=M_\mathrm{ej} / 10$~\Msun{} 
\citep[see, e.g.,][]{1987Natur.328..320S,2019ApJ...879....3G}. The macroscopic
mixing processes proceed over this time. Consequently,
the computational time is very long in order to catch all relevant dynamical
effects. 

These models were mapped into the \verb|PROMETHEUS| code in order to simulate the
contraction, bounce, shock formation, shock revival and shock propagation
until the moment of shock breakout. The details of the simulations are fully
described in the recent paper by \citet{2020MNRAS.496.2039S}, therefore, we
direct the reader to this comprehensive study for the details to avoid repetition.

In Table~\ref{table:values}, we list relevant properties of the default models. 
Furthermore, we add the subsets of models we used for our study as
explained in sections below. E.g. we did a ``Tracer''-study for the model e8.8 in which we
modified the mass of radioactive nickel $^{56}$Ni{} according to the amount
of so-called ``Tracer'' material (Section~\ref{subsect:tracer}).  We 
deem an exploration of metallicity dependence very important for the observational properties
of our models. In order to do this, we constructed a subset of models for e8.8
by tuning the iron content in the hydrogen-rich envelope, either setting
iron mass fraction to zero or to 0.00014, thus
mimicking the zero-metalicity progenitor and the progenitor at
Small Magellanic Cloud (SMC) metallicity. We did
the same experiment for model z9.6, studying three additional
metallicities: solar metallicity (set to 0.014 and 0.02 in accordance
with \citealt{2009ARA&A..47..481A} and \citealt{2003ApJ...591.1220L}) and SMC metallicity
(Section~\ref{subsect:Zdepend}). For the model e8.8, we carried out an
analysis of the influence of a different ratio between hydrogen and helium
fraction in the outer envelope, as this ratio may differ taking the
uncertainty of stellar evolution calculations into account
(Section~\ref{subsect:HHedepend}). We also carried out the radius-dependence
study for e8.8, in which we produced 3 additional models with the radius of
400~\Rsun{}, 600~\Rsun{}, and 900~\Rsun{}, while truncating the original 1D 
stellar evolution profile prior to the collapse.

In the current study, the models were mapped into the radiation hydrodynamics code \verb|STELLA|
\citep{2006A&A...453..229B}. This code is capable of processing hydrodynamics as well as radiation field evolution and
computing light curves, spectral energy distribution and resulting broad-band
magnitudes and colours. We used the standard parameter settings,
well-explained in many papers involving \verb|STELLA| simulations
\citep[see e.g., ][]{2019MNRAS.483.1211K,2020MNRAS.497.1619M}\footnote{We note that
the thermalisation parameter was set to unity in contrast to the value
of 0.9 recommended by the most recent study \citet{2020MNRAS.499.4312K}.}.

We compare our radiative transfer results to the CCSN progenitor model L15 
\citep{2000ApJS..129..625L,2017ApJ...846...37U}, which is presented in
Table~\ref{table:values} as L15-nu and L15-tb. We use this model as a ``reference'' CCSN model,
even though the progenitors of CCSNe and their explosions are diverse. 
This model approximates the explosion of a
massive progenitor with an initial mass of 15~\Msun{} at solar
metallicity, neglecting wind mass-loss \citep{2000ApJS..129..625L}. 
We note though that the published light curves \citep{2017ApJ...846...37U}
mistakenly did not account for metallicity, i.e. there is no stable iron content
in the hydrogen-rich envelope. We correct for this oversight in our calculations and
figures. Even though there is a minor effect on the bolometric light curve,
the bigger impact is observed for the \emph{U}-band magnitude and colour
temperature (see Section~\ref{subsect:Zdepend}).
The model's final mass is thus the same as the initial mass, the resulting ejecta mass is 13.5~\Msun{}, and the radius at the moment of
core-collapse explosion is 627~\Rsun{}. For comparison, we used two
types explosions for this model. The first explosion, named ``L15-nu'', was done
as a 3D neutrino-driven model with \verb|PROMETHEUS| \citep{2017ApJ...846...37U}, i.e. the \verb|PROMETHEUS| output
was mapped directly into \verb|STELLA|. The resulting light curve and
observables are labelled ``L15-nu'' in the plots below. The second
explosion, named ``L15-tb'', was
performed as a detonation of the progenitor L15 using
the thermal bomb method, assuming an explosion energy of 0.9~foe (1\,foe = $10^{\,51}$~erg)
and reaching a terminal kinetic energy of 0.54~foe. The result of these
simulations is labelled ``L15-tb'' in the plots below. The principal difference
between a 3D neutrino-driven explosion and thermal bomb explosion lies in the
absence of macroscopic mixing of the chemical composition in the latter case
mapped into \verb|STELLA|.

In Figure~\ref{figure:chemie}, we provide the input chemical profiles
for the models in our study as well as the chemical structure of L15
for comparison. Dashed lines present pre-explosion chemical structures while
solid lines present post-explosion distributions of hydrogen, helium, carbon
and oxygen. The different chemical structures of the models is easily seen in the
plot. The reference massive star model prior to the explosion exhibits a stratified chemical
structure: the outer, hydrogen-rich, envelope rests on top of a thick
2~\Msun{} pure helium shell, which in turn lies on an almost pure oxygen shell. In
contrast, the low-mass model z9.6 has a thin 0.5~\Msun{} helium layer
and no oxygen layer, and the low-mass model s9.0 has no distinct helium shell at all.
This is a result of macroscopic mixing occuring during the passage of the shock. 
At the same time the model L15 experiences macroscopic mixing
and has the sharp chemical interfaces washed out to some degree. 
The ECSN model e8.8 exhibits a unique chemical structure, 
having only a hydrogen-rich envelope, polluted by a high fraction of helium
and some amount of carbon and oxygen because of dredge-out episode, 
depending on metallicity.
The very different chemical structure of the SN ejecta of these models
leads to a variety of observational properties of the resulting SN light curves
which we discuss in detail in the following sections. 

Further, Figure~\ref{figure:density} shows the pre-SN density structure
prior to the collapse in order to illustrate the
difference in the pre-explosion density profiles.  
The layers around the final neutron star in the model e8.8 look
very unique compared to the other low-mass models (s9.0 and z9.6) and the
reference CCSN model L15. The most important difference in model e8.8 is a
very steep density gradient at the edge of the core culminating in a very tenuous hydrogen-rich envelope.
The latter condition was shown by \citet{1976Ap&SS..44..409G} to prevent
fast recombination in typical SN ejecta.
The low-mass progenitors s9.0 and z9.6 differ from the reference massive
progenitor. These models have no appreciable helium layer, while the
reference massive star explosion L15 has a relatively massive helium shell
($0.19<M_r/M_\mathrm{tot}<0.3$) prior to the explosion. Apparently, the density structure
is washed out after the passage of the shock. However, the shock propagation is
different in these three different models and unavoidably influences
a variety of properties of the ejecta and resulting observables.


\section[Results]{Results}
\label{sect:results}

\subsection[Bolometric properties]{Bolometric properties}
\label{subsect:bol}

\begin{figure}
\centering
\hspace{-5mm}\includegraphics[width=0.5\textwidth]{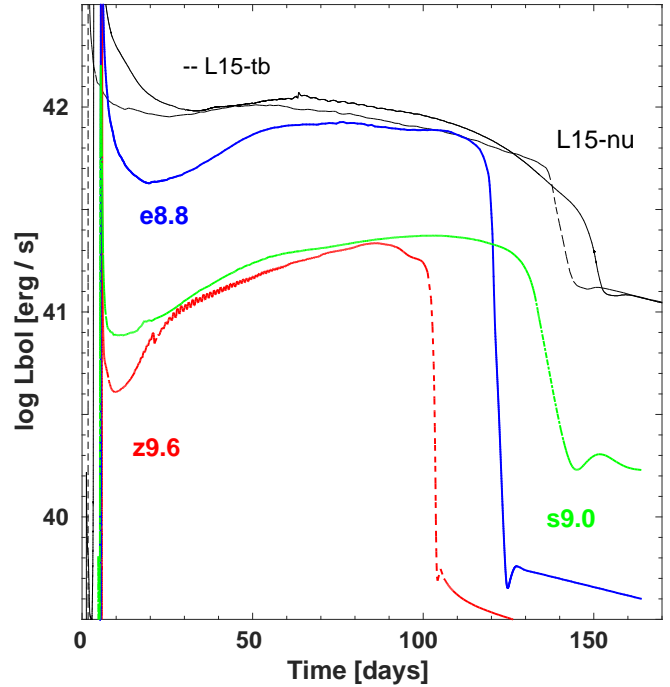}
\caption{Bolometric light curves for the models e8.8, z9.6, 
s9.0, and L15-tb/L15-nu \citep{2017ApJ...846...37U}.}
\label{figure:bol}
\end{figure}

In Figure~\ref{figure:bol}, bolometric light curves 
for the models explored in the current study are displayed.
The bolometric light curves for the models e8.8 and z9.6 clearly
differ from the canonical CCSN light curve, e.g. the curve labeled ``L15-tb'' and ``L15-nu'' 
from \citet{2017ApJ...846...37U}. The difference consists in the sharp and
pronounced drop between the plateau and the radioactive tail. The obvious
explanation is that the models in our study produce too small amount of
radioactive nickel $^{56}$Ni{}, about 0.001~\Msun{} (see Table~\ref{table:values}).
In contrast, the model s9.0 produces slightly higher amount of
$^{56}$Ni{}, about five times more (0.005~\Msun{}). Additionally
the models e8.8 and z9.6 maintain their stratified structure even at the moment
when all mixing processes cease. In contrast, there is significant large-scale
mixing in the model s9.0. Concequently, the transition from the
plateau to the radioactive tail is shallower in s9.0. The steepess of the
transition in the models e8.8 and z9.6 is also caused by a lack
of a distinct oxygen shell 
\citep[Figure~\ref{figure:chemie}, ][]{2013ApJ...772..150J,2020MNRAS.496.2039S}.
As seen, the plateau luminosities for the low-mass CCSN models z9.6 and s9.0
are lower than that of a canonical CCSN originating from a standard mass
progenitor\footnote{The light curves L15-nu and L15-tb perform explosion with
almost the same terminal kinetic energy of 0.54/0.55~foe, i.e. the plateau
luminosity is expected to be the same. However, the light curve ``L15-nu'' has
higher luminosity due to the additional heating from the extended mixing of
radioactive {}$^{56}$Ni which is responsible for an extra energy budget
\citep{2019MNRAS.483.1211K}.}, this is due to the relatively lower explosion energy. Nevertheless,
the plateau luminostity 
for the ECSN model e8.8 is comparable to the normal CCSN as a result of the
large radius of the progenitor ($L_\mathrm{bol} \sim R^{\,0.76}$, \citealt{2019ApJ...879....3G}).

\begin{figure}
\centering
\large\bf\textsf{\input{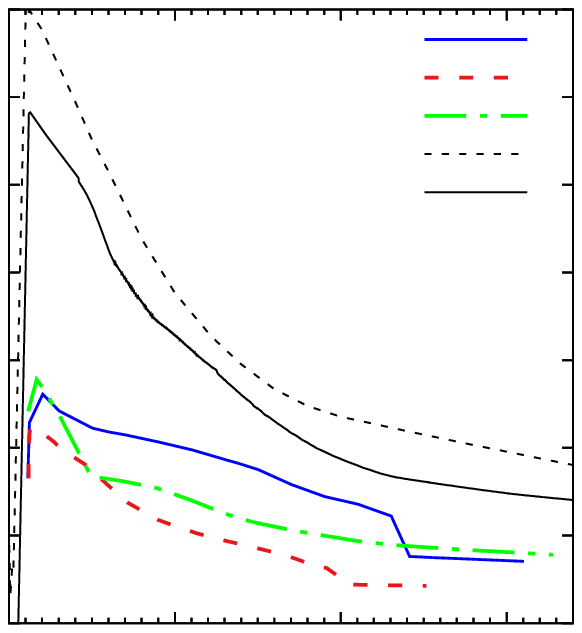}}
\caption{Time evolution of the photospheric velocity, U$_\mathrm{ph}$, for the models e8.8, z9.6, 
s9.0, and L15-tb/L15-nu.}
\label{figure:uph}
\end{figure}

In order to illustrate the overall energetics of the explosions,
we show the photospheric velocity evolution for the models considered here together
with the SN~IIP reference models L15-nu and L15-tb in Figure~\ref{figure:uph}. The
photospheric velocity is the velocity of the mass shell in which the accumulated
optical depth in \emph{B} broad band is equal 2/3.
It is easily seen that the photospheric velocity in all of our models, i.e.
both the ECSN model and
low-mass CCSN models, is systematically lower than the photospheric velocity for
the reference SN~IIP models L15-nu/L15-tb because of the lower explosion energy.
Indeed photospheric velocity at the earliest phase is already very low,
2500~km\,s$^{\,-1}$, while the reference CCSN model exhibits velocity of
about 7000~km\,s$^{\,-1}$, typical for observed SNe~IIP \citep{2009ApJ...696.1176J}. 
The photospheric velocity remains steady at a level of 1000--2000~km\,s$^{\,-1}$
thoughout the entire plateau in all our models. However, the ECSN model
exhibits roughly 500~km\,s$^{\,-1}${} higher velocity during the plateau,
i.e. slightly faster ejecta compared to the low-mass CCSN models in our
study. This is explained by the fact of relatively lower ejecta mass.
Velocity scales with energy $v\sim\sqrt{E/M}$, i.e. an increase a magnitude
in energy (expected explosion energy for a ECSN is
$10^{\,50}$~erg, while typical explosion energy of an CCSN is $10^{\,51}$~erg) 
leads to an increase in velocity about
three times over, as seen in the plot. We suggest to use this diagnostic as
a distinct feature in the identification of low-mass explosions, both ECSNe and
low-mass CCSNe, which is in agreement with the suggestions by
\citet{2017MNRAS.464.3013P} and \citet{2018MNRAS.475.1937T}.

\subsection[Broad-band light curves]{Broad-band light curves and colours}
\label{subsect:bands}

\begin{figure}
\centering
\includegraphics[width=0.5\textwidth]{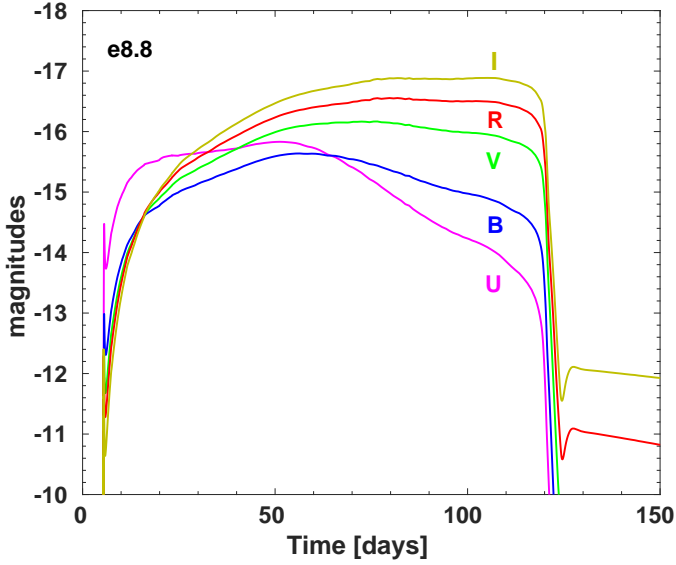}
\caption{Broad-band \emph{U}, \emph{B}, \emph{V}, \emph{R}, and \emph{I} 
light curves for the model e8.8.}
\label{figure:e88bands}
\end{figure}

\begin{figure}
\centering
\includegraphics[width=0.5\textwidth]{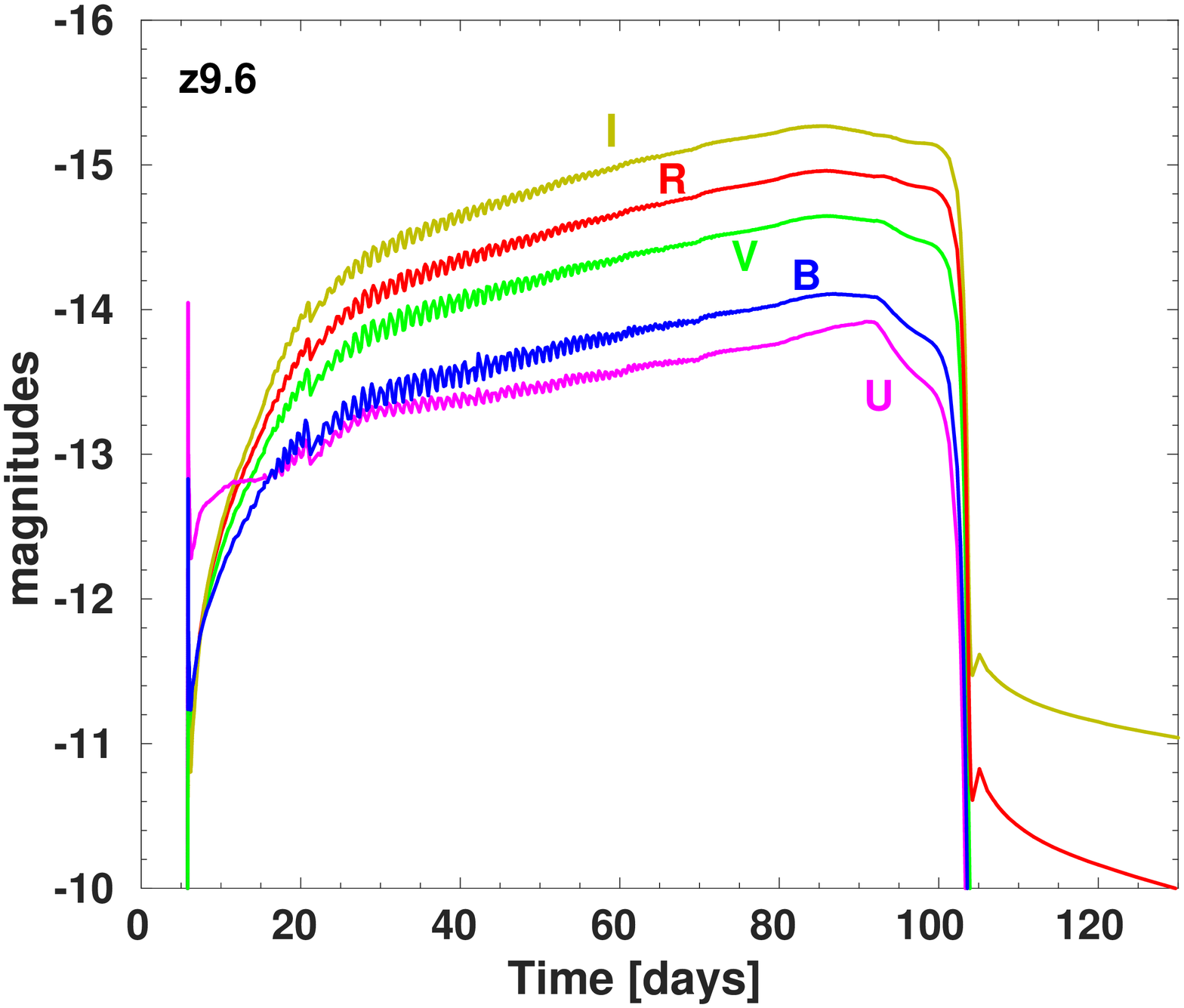}
\caption{Broad-band \emph{U}, \emph{B}, \emph{V}, \emph{R}, and \emph{I} 
light curves for the model z9.6.}
\label{figure:z96bands}
\end{figure}

\begin{figure}
\centering
\includegraphics[width=0.5\textwidth]{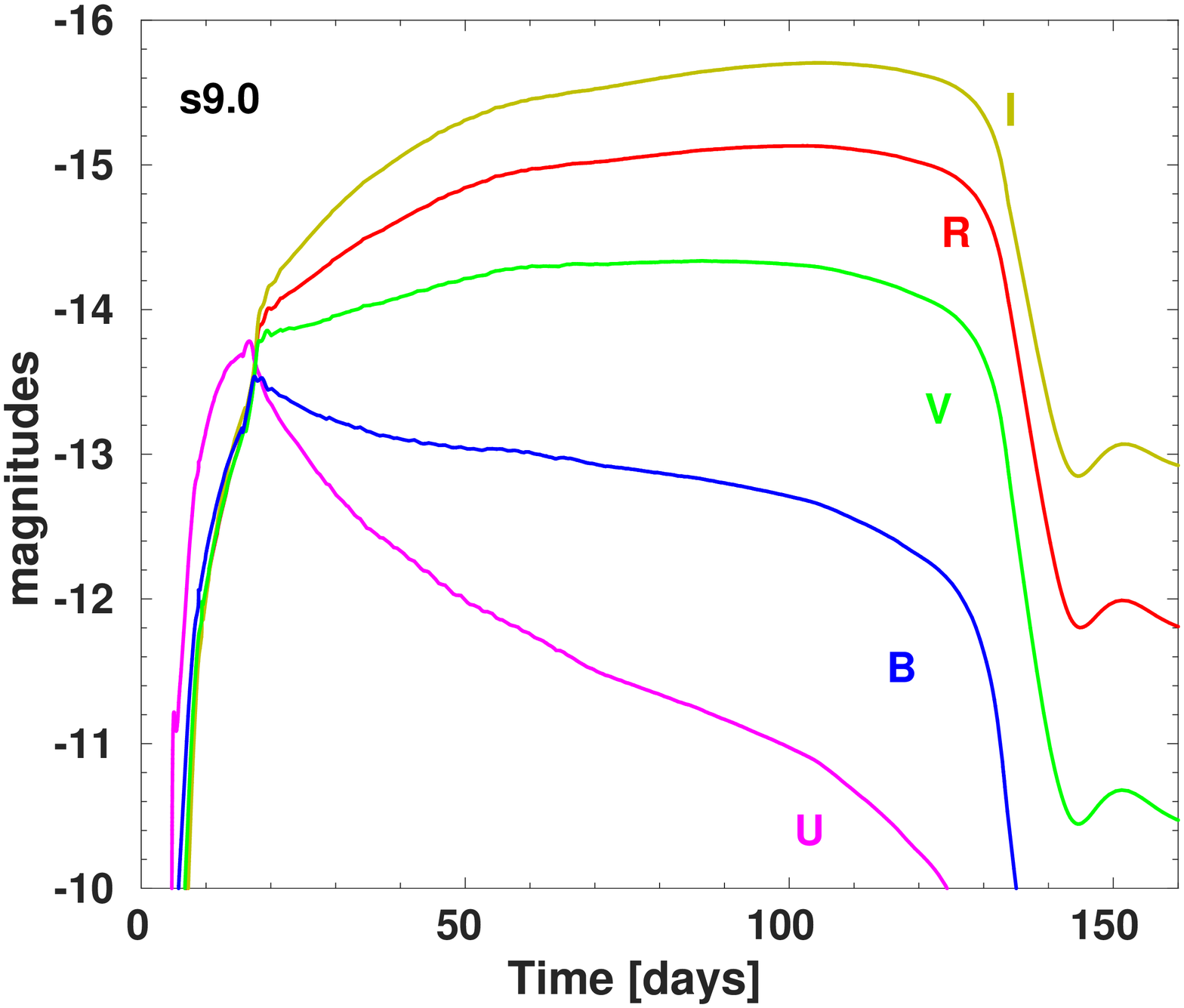}
\caption{Broad-band \emph{U}, \emph{B}, \emph{V}, \emph{R}, and \emph{I} 
light curves for the model s9.0.}
\label{figure:s90bands}
\end{figure}

We show the light curves in the standard Bessel broad-bands for the models
e8.8, z9.6, and s9.0 in
Figures~\ref{figure:e88bands}, \ref{figure:z96bands}, and
\ref{figure:s90bands}, respectively.

The broad-band light curves for the model e8.8 look similar to broad-band
magnitudes for a reference SN~IIP, excepting the first 50~days. 
The distinguishing feature of the model e8.8 is a 50-day blue plateau, i.e. the \emph{U}
band magnitude remains constant and even slightly increases over the first
half of the plateau (see Figure~\ref{figure:e88bands}), while normal SNe\,IIP
usually exhibit a decreasing \emph{U}-band
magnitude across the entire plateau. We directly compare the \emph{U} band
magnitudes of the ECSN model and the CCSN model in the next section.
We note that the original progenitor's metallicity has a strong impact 
on the behaviour of the light curve in the \emph{U}
band, specifically, the higher metallicity leads to overall redder light curves \citep[see
e.g., ][]{2000ApJ...530..966L}. 
We explore the metallicity dependence in the Section~\ref{subsect:Zdepend} below.
However, the model e8.8 shows blue colours even at solar metallicity. 
Reasons for this are the colour temperature evolution and the
conditions during, and the time of, start of recombination 
\citep{1971Ap&SS..10....3G,2014arXiv1404.6313G,2017ApJ...838..130S}. 
In contrary to normal CCSNe, recombination
settles in at relatively late times depending on the explosion energy \citep{Shussman2016,2020MNRAS.494.3927K}: 
\begin{equation}
t_\mathrm{rec}\sim M^{\,0.22} R^{\,0.76} E^{\,-0.43}\,,
\label{equation:Trecomb}
\end{equation}
where $M$ is ejecta mass in 15~\Msun{}, $R$ is progenitor radius in
500~\Rsun{}, and $E$ is explosion energy in 1~foe.
The ejecta mass for ECSN progenitors is generally expected to be lower than that of
CCSNe (e.g. for both our low-mass and the reference models). 
The explosion energy is also expected to be lower 
\citep[see e.g., ][]{2015ApJ...801L..24M,2017ApJ...850...43R,2020MNRAS.496.2039S}.
The progenitor radius of the ECSN e8.8 1200~\Rsun{} is significantly larger than 
the 214~\Rsun{} of model z9.6 (the latter being zero metallicity) and
the 408~\Rsun{} of model s9.0. Note that the e8.8 radius is still in
the range of reference progenitor radii of red
supergiants \citep[100\,--\,2850~\Rsun{}, ][]{2005ApJ...628..973L,2006ApJ...645.1102L}.
The larger radius of the ECSN progenitor (the super-AGB star), is a
consequence the dredge-out episode \citep{1987ApJ...322..206N,1999ApJ...515..381R,2013ApJ...772..150J}.
Accordingly, the hydrogen-rich envelope of the model e8.8 is very tenuous and
recombination hard settles in, which in turn forces the overall
electron-scattering photosphere to recede more slowly \citep{1976Ap&SS..44..409G}.

The zero-metallicity model z9.6 demonstrates a quite unusual behaviour in the broad-band light curves
compared to a normal SN~IIP. 
We show in the next sections that the same model
at solar metallicity should have colour behaviour quite standard for SNe~IIP.
Nevertheless, the zero-metallicity low-mass CCSNe, as demonstrated by the
model z9.6, have monotonically rising \emph{U}, \emph{B}, \emph{V},
\emph{R}, and \emph{I} light curves, and constant colours during the plateau
phase. This means that the shape of the spectrum persists more or less
unchanged for the first 100~days, i.e. while the electron-scattering
photosphere gradually recedes through the hydrogen-rich envelope. Later,
i.e. after the end of plateau, there is a
sharp reddening of the colours. The photosphere at this phase enters the
inner region of the ejecta dominated by iron-group elements which are the major
contributors to the line opacity. However, compatible zero-metallicity stars
(Population~III) exist only in the early Universe, therefore, the
application of model 9.6 is of limited usefulness, when observations in the
nearby Universe are concerned. Predictions about
observational properties of CCSNe for Population~III stars are very useful for 
upcoming transient surveys like LSST\footnote{The Large Synoptic Survey Telescope
which is the core of the Vera~C.~Rubin Observatory.} \citep{2013ApJ...768...95W,2019ApJ...873..111I}. 
However, the majority of known SNe~IIP are found in non-zero metallicity
galaxies \citep{2016A&A...589A.110A}, and theoretical predictions for the
observables of CCSNe at the SMC and solar metalicity are deemed more useful.

\subsection[Dependence on the explosion energy]{Dependence on the explosion energy}
\label{subsect:Edepend}

\begin{figure*}
\centering
\vspace{1mm}
\includegraphics[width=0.5\textwidth]{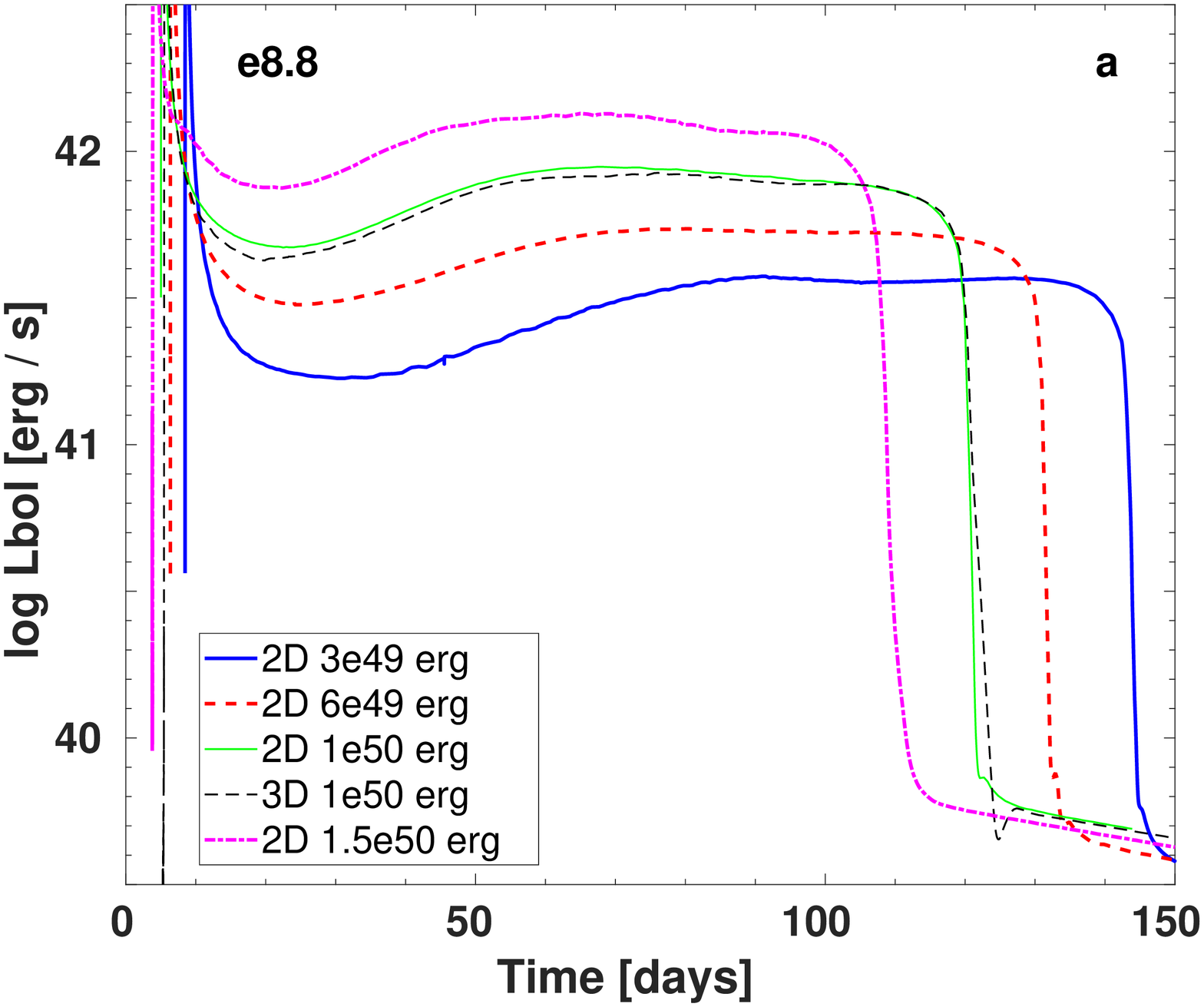}~
\hspace{1mm}\includegraphics[width=0.5\textwidth]{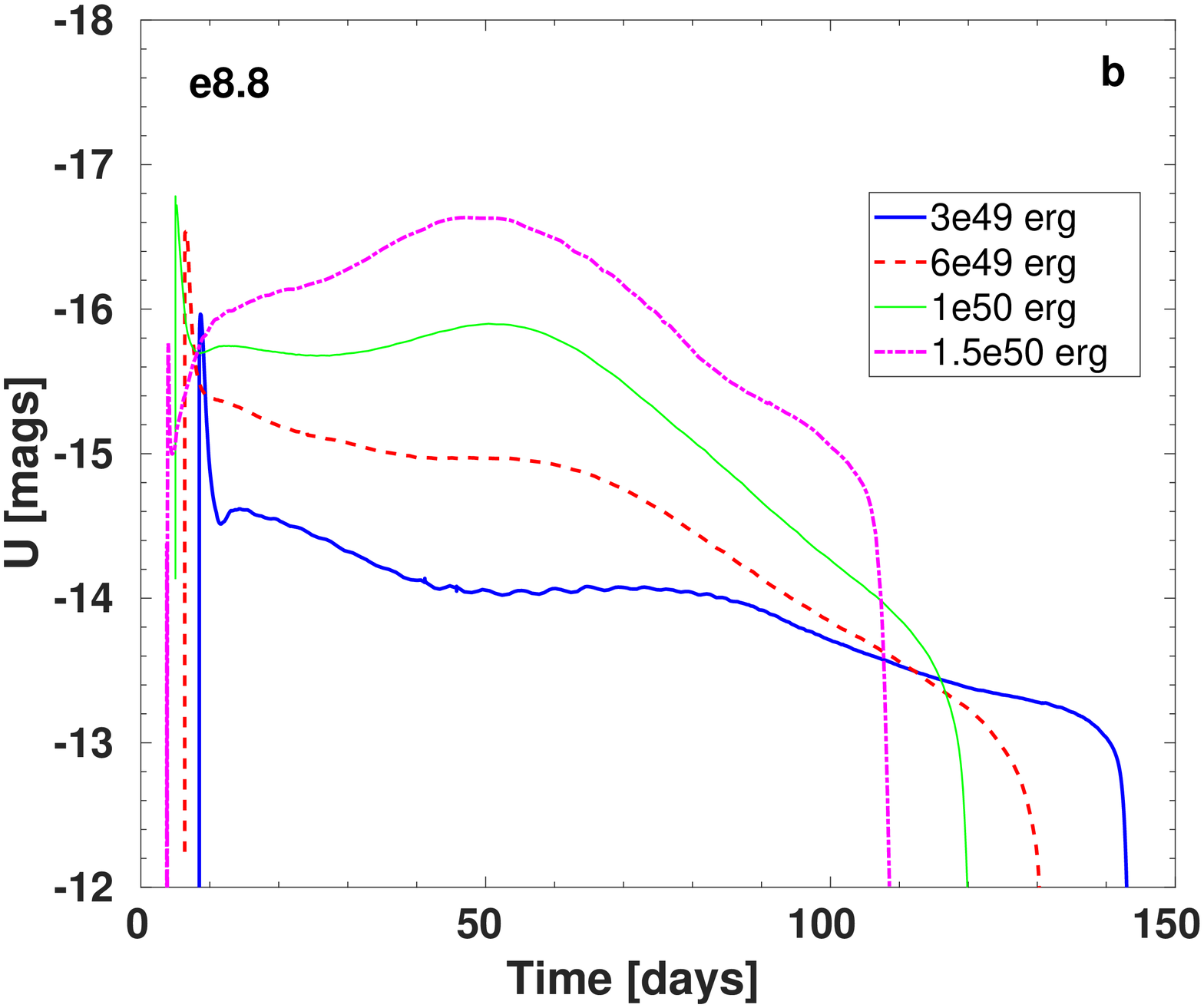}\\
\vspace{4mm}
\includegraphics[width=0.5\textwidth]{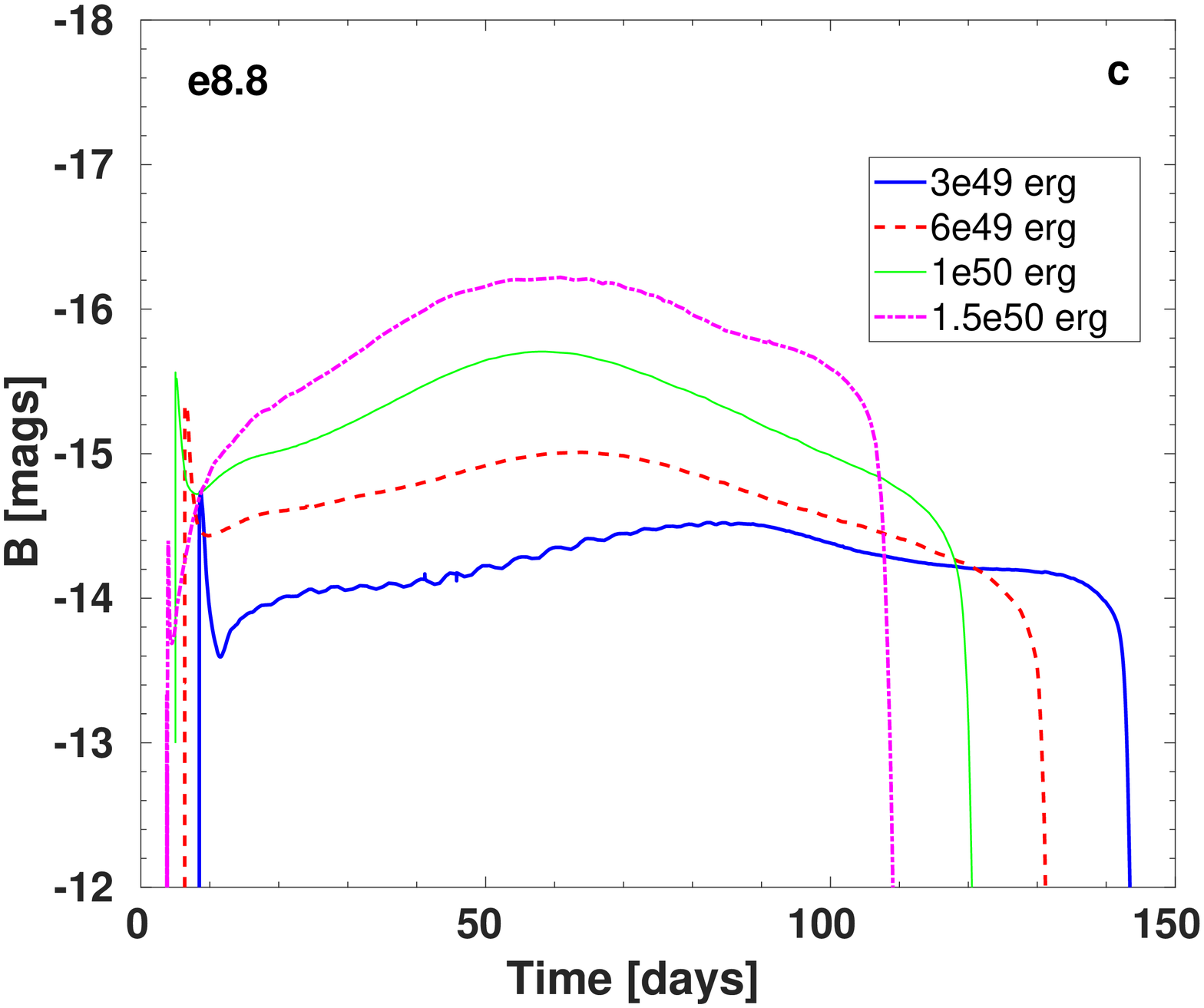}~
\hspace{1mm}\includegraphics[width=0.5\textwidth]{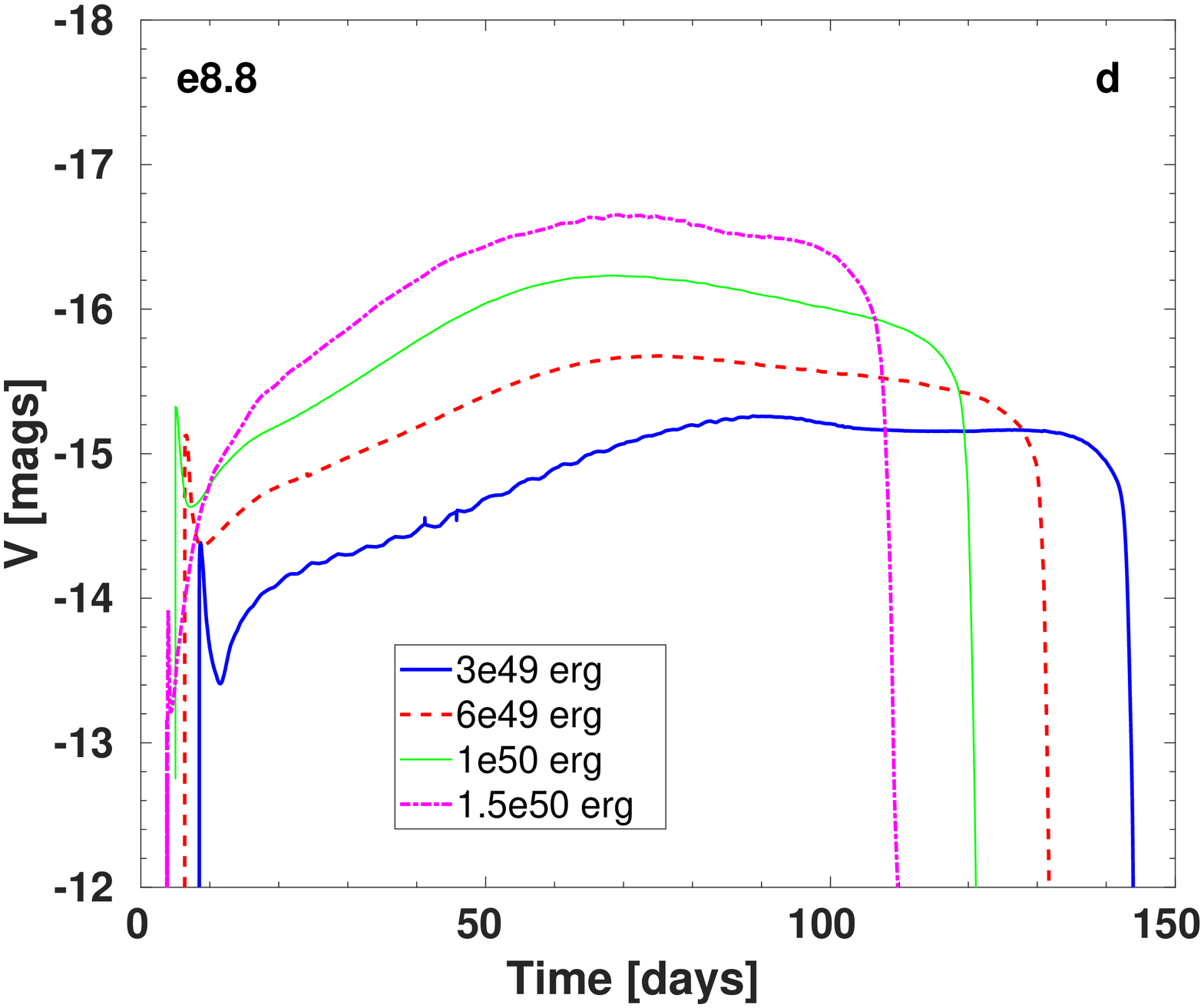}\\
\vspace{5mm}
\includegraphics[width=0.5\textwidth]{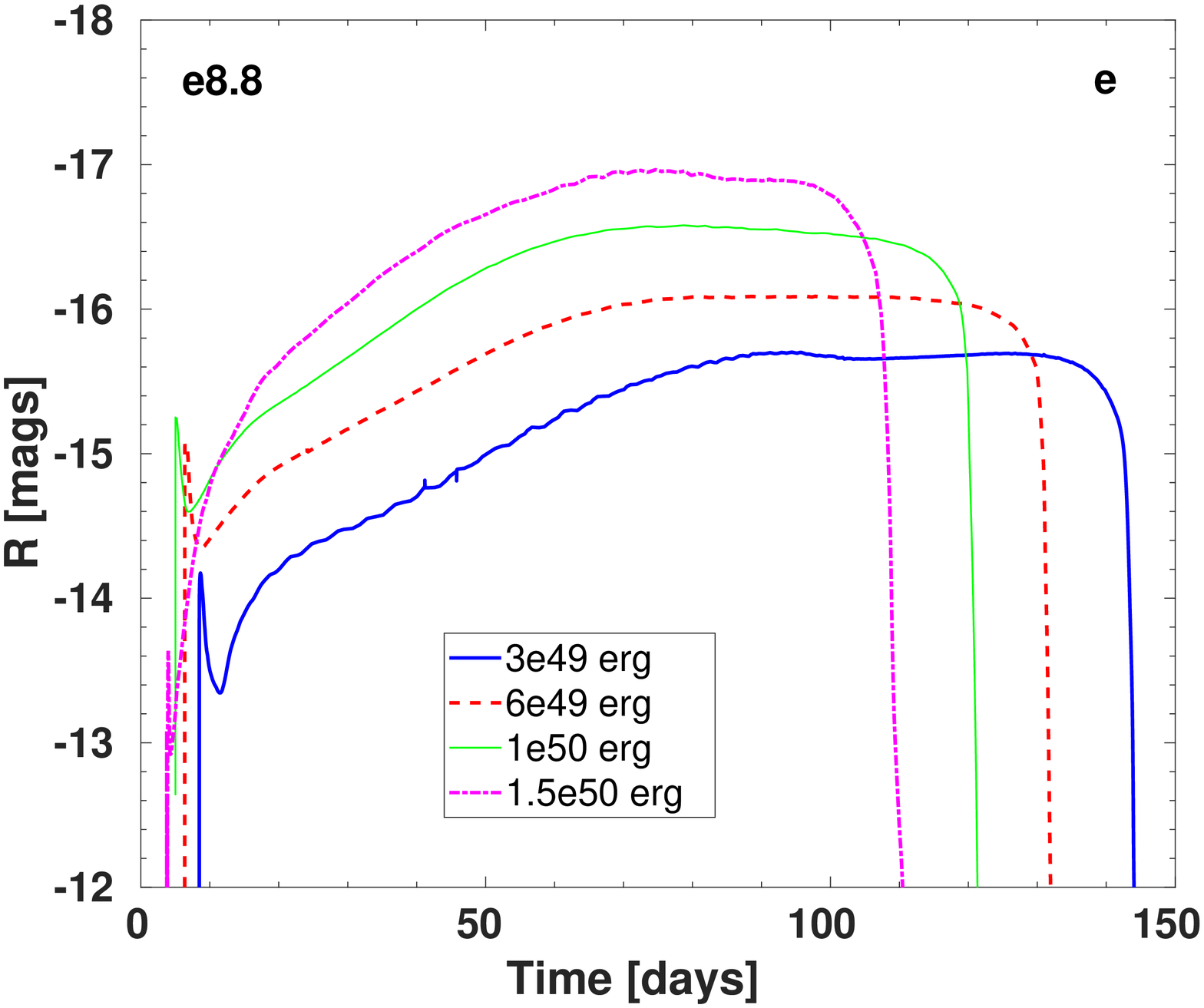}~
\hspace{4mm}\includegraphics[width=0.49\textwidth]{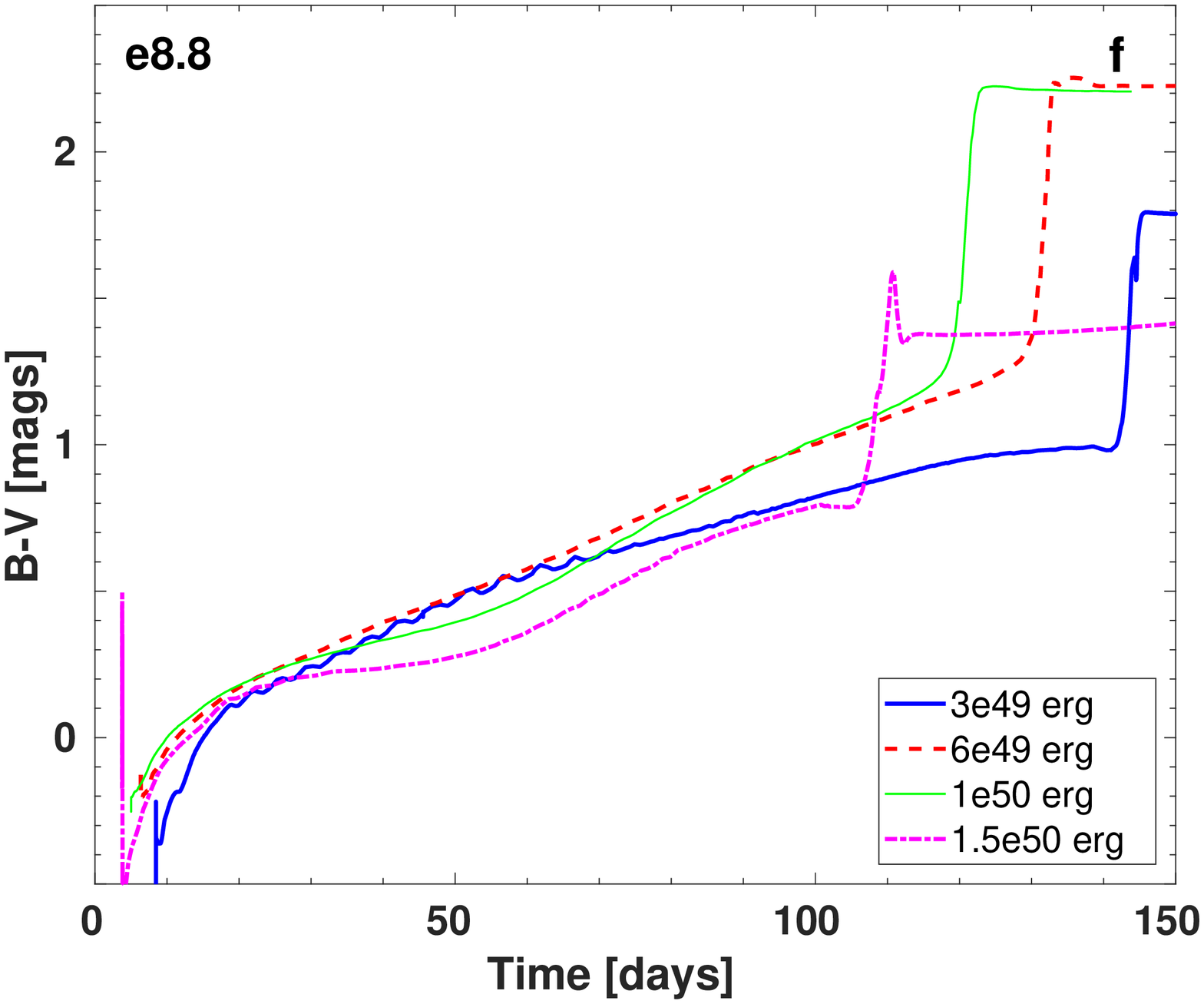}\\
\caption{Dependence on the explosion energy: 
Bolometric, broad-band light curves, and \emph{B}--\emph{V}
colour for the model e8.8 simulated in 2D and 3D with
explosion energies as labelled.}
\label{figure:Edepend}
\end{figure*}

The numerics and physics input in the core-collapse simulations lead 
to uncertainties in the final
explosion energy \citep{2017ApJ...850...43R,2020ApJ...891...27M}.
As a remedy, \citet{2020MNRAS.496.2039S} provided simulations for 
e8.8 with four different neutrino luminosity values in 2D
corresponding to explosion energies of: $3\times10^{\,49}$~erg, $6\times10^{\,49}$~erg,
$1\times10^{\,50}$~erg, and $1.5\times10^{\,50}$~erg (see Table~\ref{table:values}).
We would like to emphasize that the explosion energy is unequal to the terminal kinetic
energy listed in Table~\ref{table:values}. There are number of reasons for this
difference. (1) The explosion energy published by \citet{2020MNRAS.496.2039S}
is a direct integral of the total energy in 2D or 3D.
The multidimensional profiles were converted into 1D-profiles using an
angle-averaging procedure. (2) Since \verb|STELLA| is a hydrodynamics
code, it requires some numerical relaxation when mapping \verb|PROMETHEUS| output into it, 
which in turn is liable to cause some (up to 15\,\%) difference in the resulting integrated
energy. (3) The supernova ejecta at the moment of shock breakout have not
yet reached the coasting phase. 
This causes some hydrodynamical evolution and inelastic conversion of
kinetic energy into thermal energy.
We note that the total mass of radioactive nickel $^{56}$Ni{} is kept the
same for different explosion energies. Nevertheless, it is expected that
more energetic explosions naturally produce a higher mass of $^{56}$Ni{}
\citep{2016ApJ...818..124E}. However, we find that
the resulting chemical profiles in the post-shock ejecta structure do not
differ significantly. For different energies the hydrodynamical profiles explicitly
scale with the explosion energy. Otherwise, the final light curves obey the
well-known Popov's relation for absolute magnitude in \emph{V} band and
duration of the plateau $t_p$
\citep{1985SvAL...11..145L,1993ApJ...414..712P,2016ApJ...821...38S,2019ApJ...879....3G}:
\begin{equation}
\begin{aligned}
V \sim - 1.67 \log R + 1.25 \log M - 2.08 \log E \\
\log t_p \sim 0.167 \log R + 0.5 \log M - 0.167 \log E\,,
\end{aligned}
\label{equation:Popov}
\end{equation}
where $R$ is radius in \Rsun{}, $M$ is ejecta mass in \Msun{}, and $E$ is
explosion energy in foe.
In Figure~\ref{figure:Edepend}, bolometric and broad-band light curves and
\emph{B}--\emph{V} colour for the model e8.8 computed with a variety of energies
are presented. Figure~\ref{figure:Edepend}a also shows the
result of the simulations for the model e8.8 in 3D for an explosion energy
of $1\times10^{\,50}$~erg (bolometric light curve for the default model e8.8
is displayed in Figure~\ref{figure:bol}). The light curves for this case are
almost identical in 2D and 3D. This means that an explosion calculated in 2D provides the
same hydrodynamical and chemical ejecta structure as in 3D. The higher explosion energy leads to
a more luminous and
shorter plateau. The broad-band magnitudes for this case follow the
behaviour of the bolometric curve
without major flux redistribution thoughout the spectrum. Figure~\ref{figure:Edepend}
shows \emph{B}--\emph{V} colour for different
choices in explsoion energy. There is no significant difference for cases of
different energy. However, the colour tends to be slightly bluer for higher
energies, e.g. \emph{B}--\emph{V} is 0.2~mag bluer an explosion
energy of $1\times10^{\,50}$~erg and $1.5\times10^{\,50}$~erg in e8.8
at the end of plateau and later compared to the results for the lower explosion energy of $3\times10^{\,49}$~erg
and $6\times10^{\,49}$~erg.

\subsection[Dependence on the initial metallicity]{Dependence on the initial metallicity}
\label{subsect:Zdepend}

\begin{figure*}
\centering
\hspace{4mm}\includegraphics[width=0.49\textwidth]{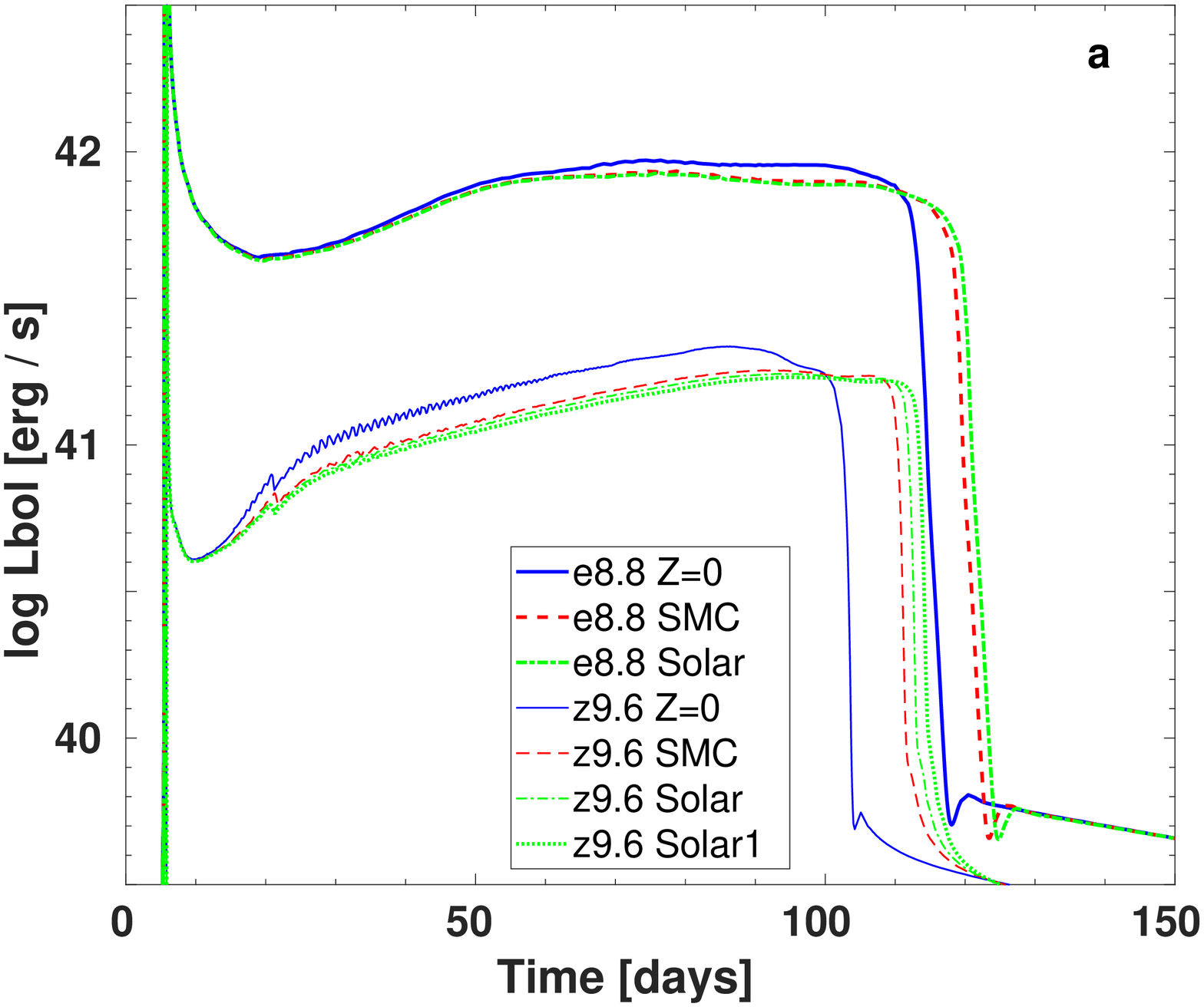}~
\includegraphics[width=0.5\textwidth]{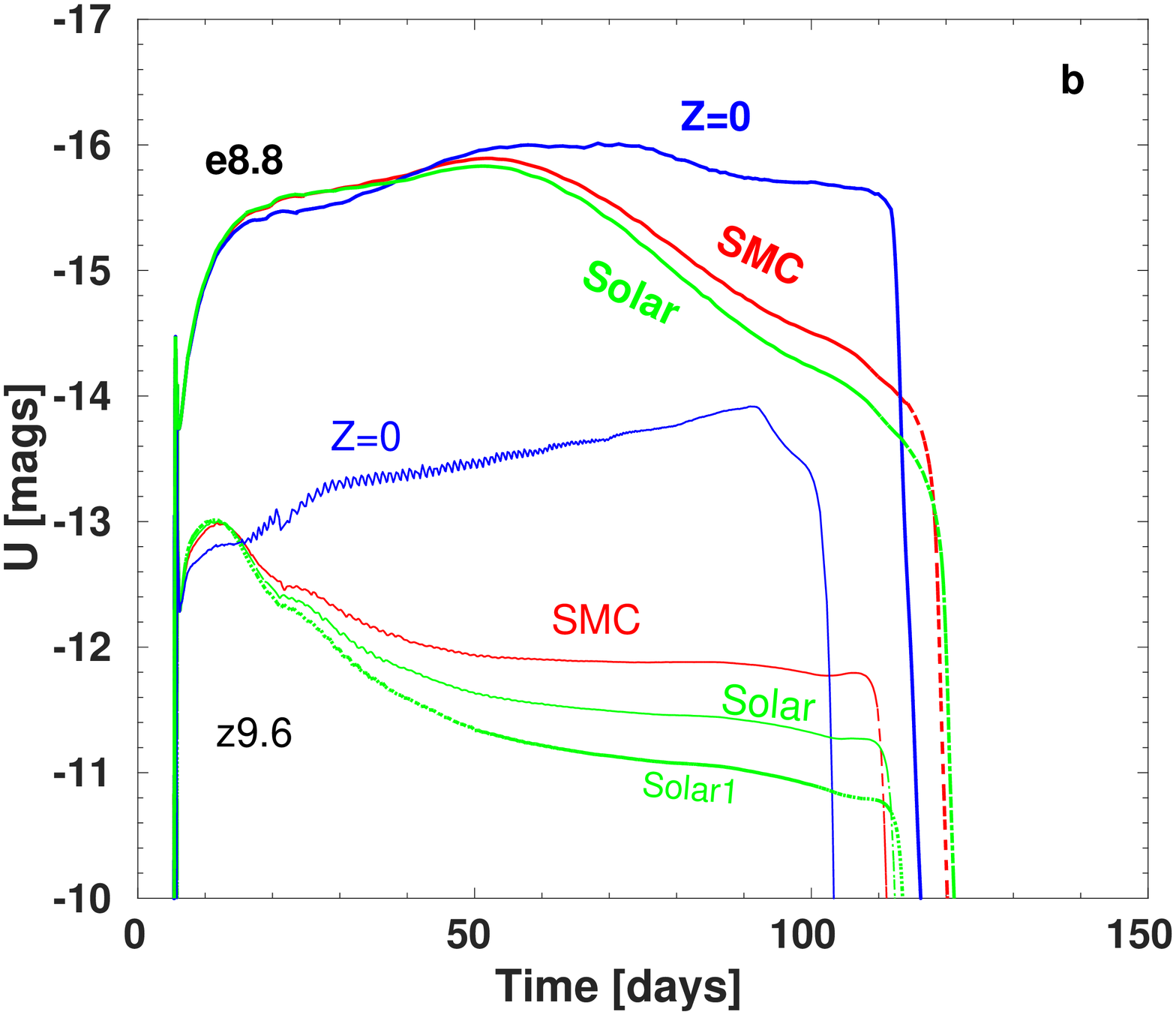}\\
\includegraphics[width=0.5\textwidth]{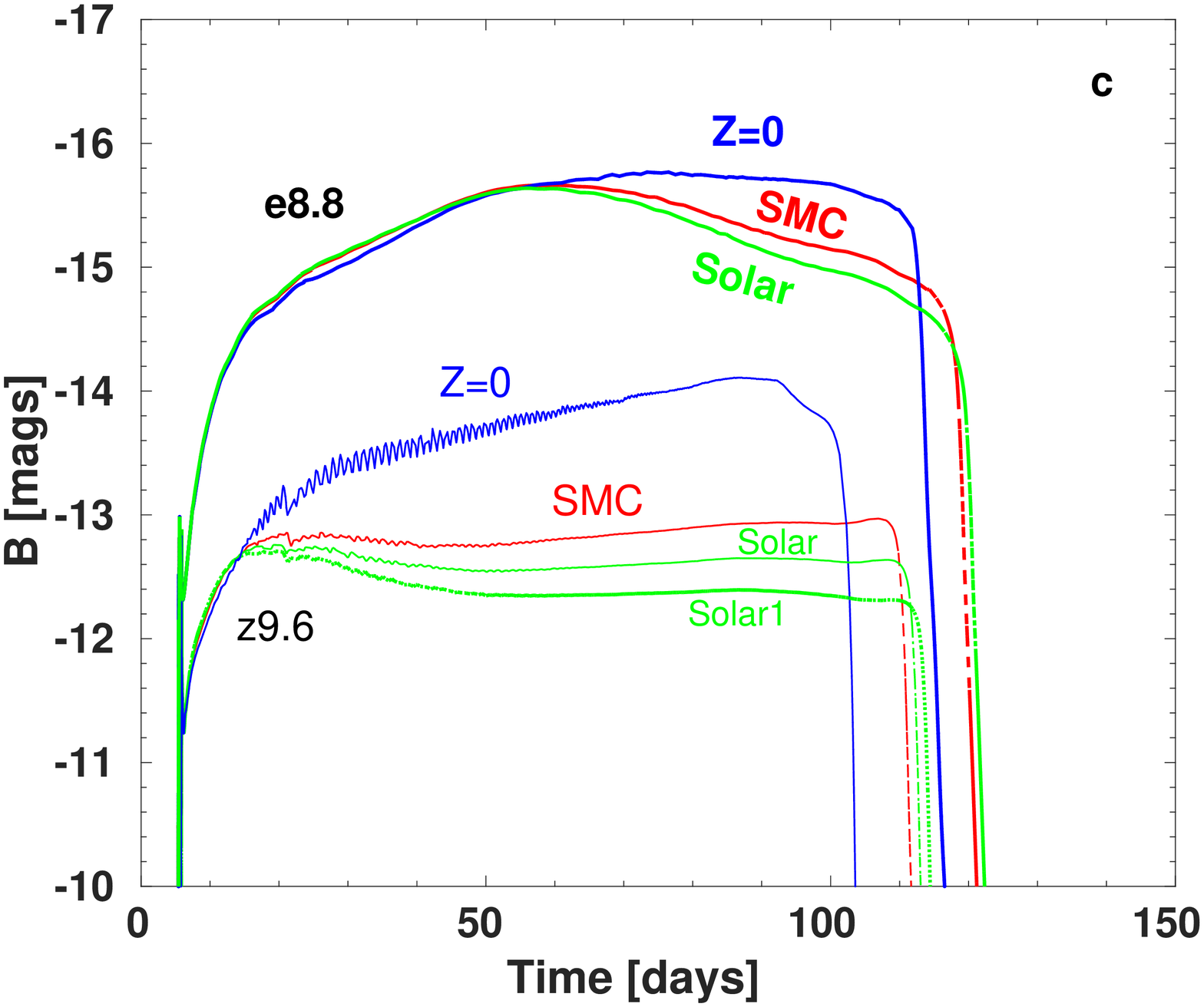}~
\includegraphics[width=0.5\textwidth]{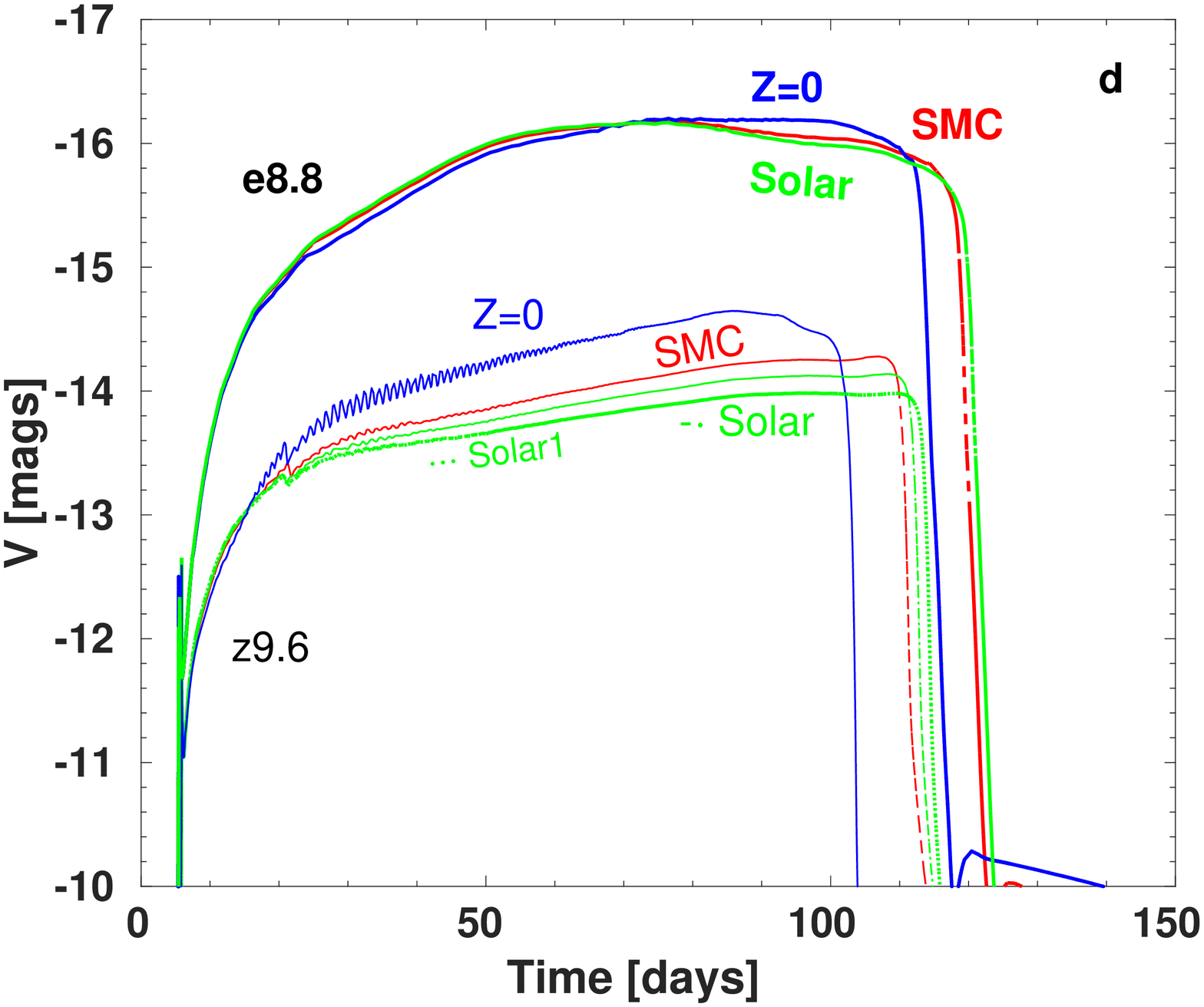}\\
\includegraphics[width=0.5\textwidth]{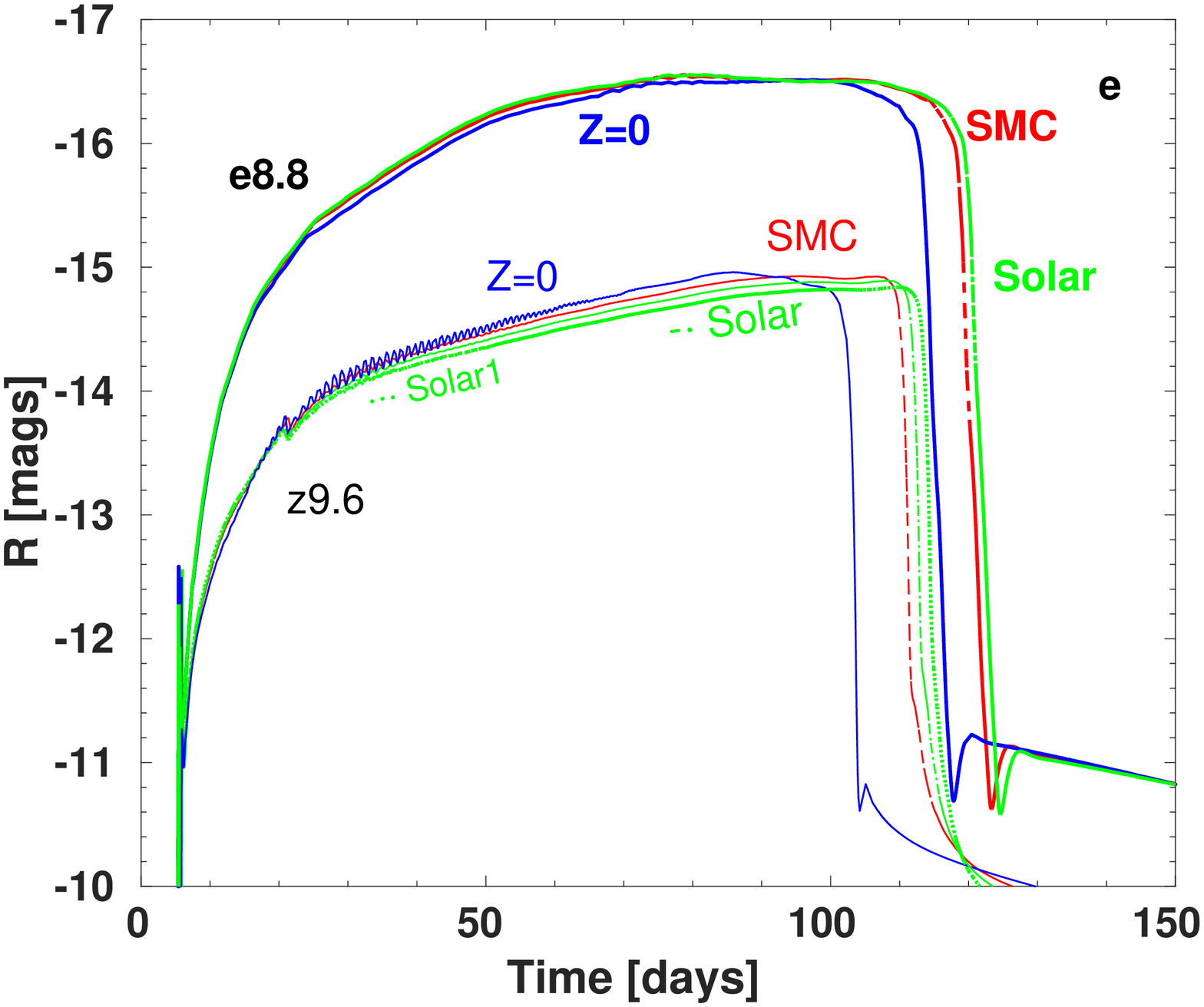}~
\hspace{5mm}\includegraphics[width=0.485\textwidth]{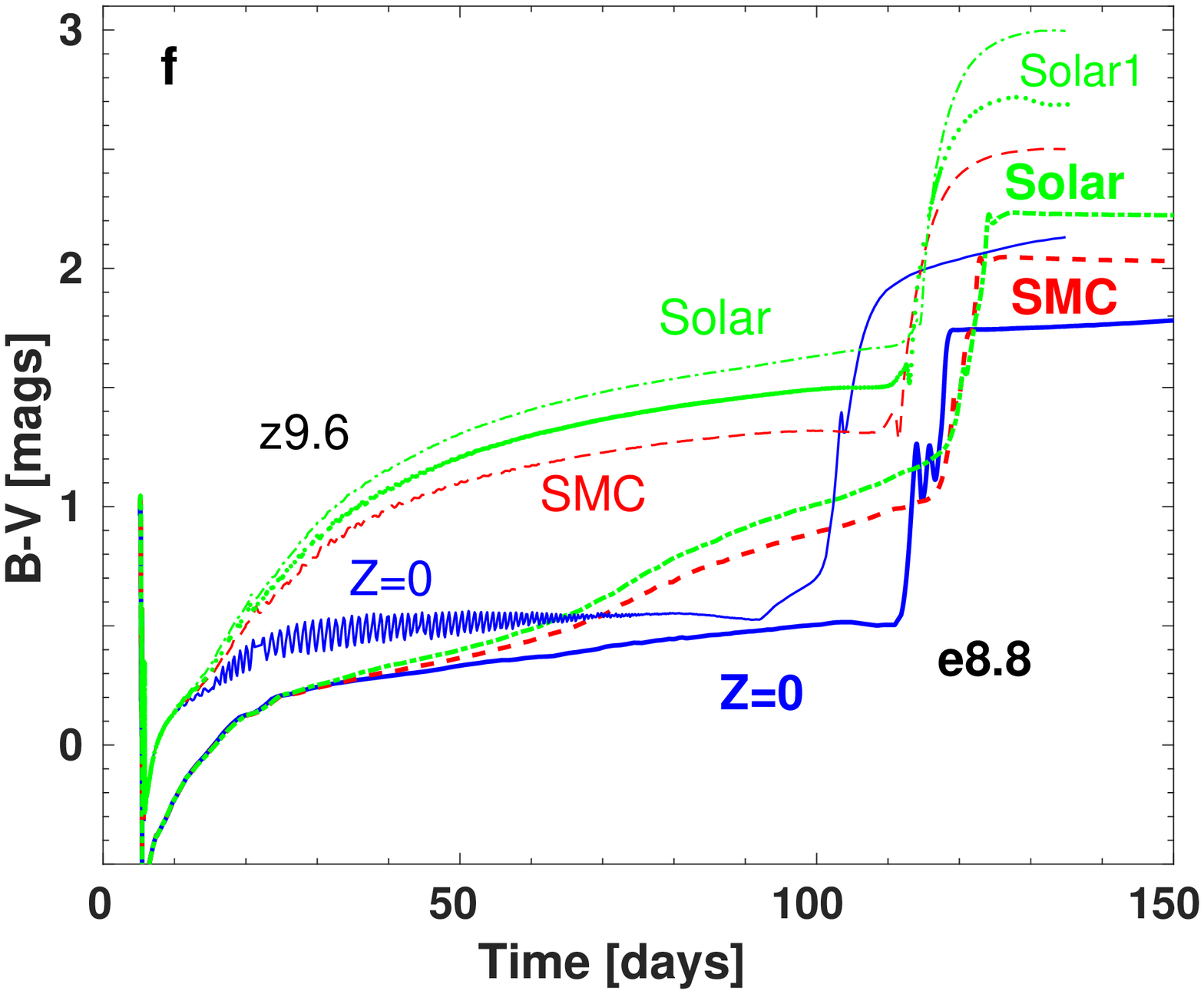}\\
\caption{Dependence on the initial metallicity: Bolometric and broad-band light curves, and \emph{B}--\emph{V} 
colour for the subset of runs based on the model e8.8
and z9.6 with different stable-Iron content in the hydrogen-rich envelope,
which corresponds to zero metallicity (``Z=0''), SMC metallicity (``SMC''),
and solar metallicity (``Solar'' and ``Solar1''). ``Solar'' metallicity
stands for the initial metallicity of the model e8.8 and has iron fraction
of $5\times10^{\,-4}$, while ``Solar1'' metallicity
stands for the solar metallicity with the iron fraction of
$1.4\times10^{\,-3}$ (initial metallicity of the model s9.0).}
\label{figure:Zdepend}
\end{figure*}

In order to understand the influence of initial metallicity, 
we explored two of the models from the study in greater detail: e8.8 and z9.6. Specifically, we modified
the stable iron abundance in the hydrogen-rich envelope. 
The subset of the runs for model e8.8 has the following adopted metallicity: 
zero ($Z=0$), SMC ($Z=0.0014$), solar ($X(\mathrm{Fe})=5\times
10^{\,-4}$), the latter being the default metallicity of the model e8.8).
The subset of the runs for the model z9.6 has the adopted metallicity: zero
($Z=0$, the default metallicity of the input model), SMC ($Z=0.0014$,
$X(\mathrm{Fe})=1.4\times10^{\,-4}$),
and two different runs for solar metallicity ``solar'' ($Z=0.014$,
\citealt{2009ARA&A..47..481A}, $X(\mathrm{Fe})=5\times 10^{\,-4}$, the default metallicity of the model e8.8) and
``solar1'' ($Z=0.02$, \citealt{2003ApJ...591.1220L}, $X(\mathrm{Fe})=1.4\times
10^{\,-3}$, the default metallicity of the model s9.0). Changing
metallicity of the initial model unavoidably leads to different stellar
evolution path, different mass-loss, consequently different final mass and
progenitor radius \citep{2012A&A...538L...8G,2013EAS....60...43G,2015MNRAS.447.3115J,2017A&A...603A.118R}.
We, nevertheless, rely on the conclusions we make based on the
metallicity study we carried out in the frame of the current paper.

Iron is the most influencial element contributing to the overall
line opacity. Therefore, light curves for runs with zero metallicity
are significantly more blue than light curves for solar metallicity runs
as seen in Figure~\ref{figure:Zdepend} \citep[see also][]{2020Goldshtein}. The
most significantly affected broad band magnitudes are \emph{U} and \emph{B}, while
the \emph{V}, \emph{R}, and \emph{I} magnitudes are less dependent on the iron
content. This is due to the blue flux being effectively absorbed by iron
and redistributed to longer wavelengths
\citep[][]{1999A&A...345..211L,2006ApJ...649..939K,2020MNRAS.499.4312K}. 
Hence, a supernova may have a prominent
110-day plateau in \emph{U} and \emph{B} broad
band in case of zero metallicity (e8.8 at $Z=0$), while having a 50~day plateau
with a subsequent decline in case of solar and SMC metallicity (e8.8 ``solar'' and
e8.8 ``SMC''). Interestingly, the \emph{U} broad band light curve rises for the model z9.6
with the default zero metallicity, while it declines for solar/solar1 and SMC
metallicity, similar to many typical SNe~IIP. Similarly, the light
curve rises in \emph{B} band in case of zero metallicity, while the model z9.6 shows a
plateau in \emph{B} if the iron content in the hydrogen-rich envelope
corresponds to solar or SMC metallicity. Hence, the \emph{U} band light curve
serves as a direct indicator of the initial metallicity of the progenitor
for CCSNe \citep{2013MNRAS.433.1745D,2014MNRAS.440.1856D}. The same
is true for the ECSN model e8.8. However, the earlier 50-day \emph{U} band light curve
remains unaffected by metallicity, because of the relatively long delay
until recombination settles in (see Figure~\ref{figure:e88bands}).
Additionally, metallicity, i.e.
iron content in the hydrogen-rich envelope, effectively determines the length
of the plateau. Specifically, a difference of {}5 to 10~days is observed in
duration of the plateau between runs with zero and solar/SMC metallicity:
the higher metallicity the longer the plateau \cite[see e.g., ][]{2009ApJ...703.2205K}.

\subsection[Dependence on the H-to-He-ratio]{Dependence on the H-to-He-ratio in the outer envelope}
\label{subsect:HHedepend}

\begin{figure*}
\centering

\hspace{1cm}\includegraphics[width=0.49\textwidth]{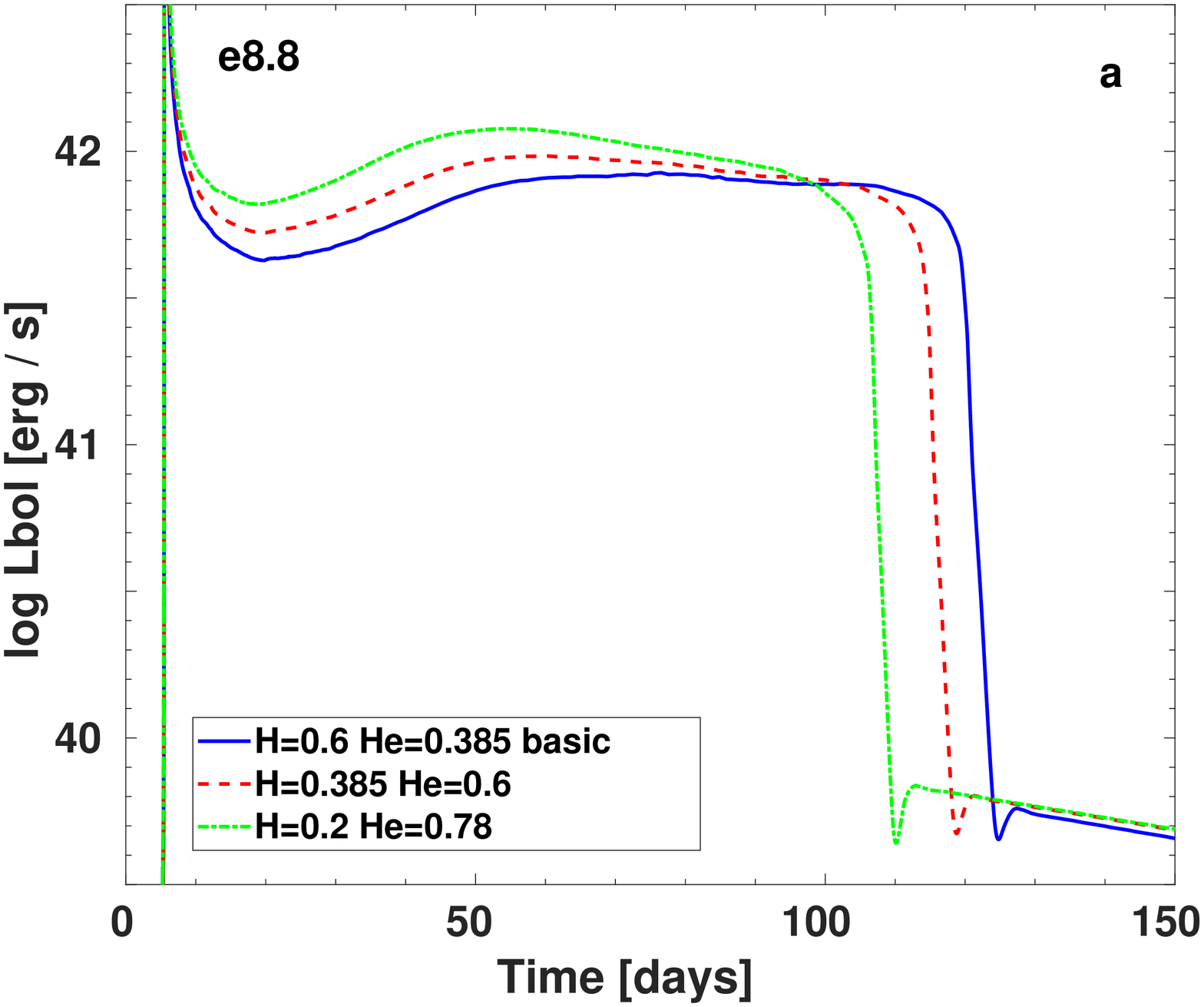}~
\hspace{2mm}\includegraphics[width=0.5\textwidth]{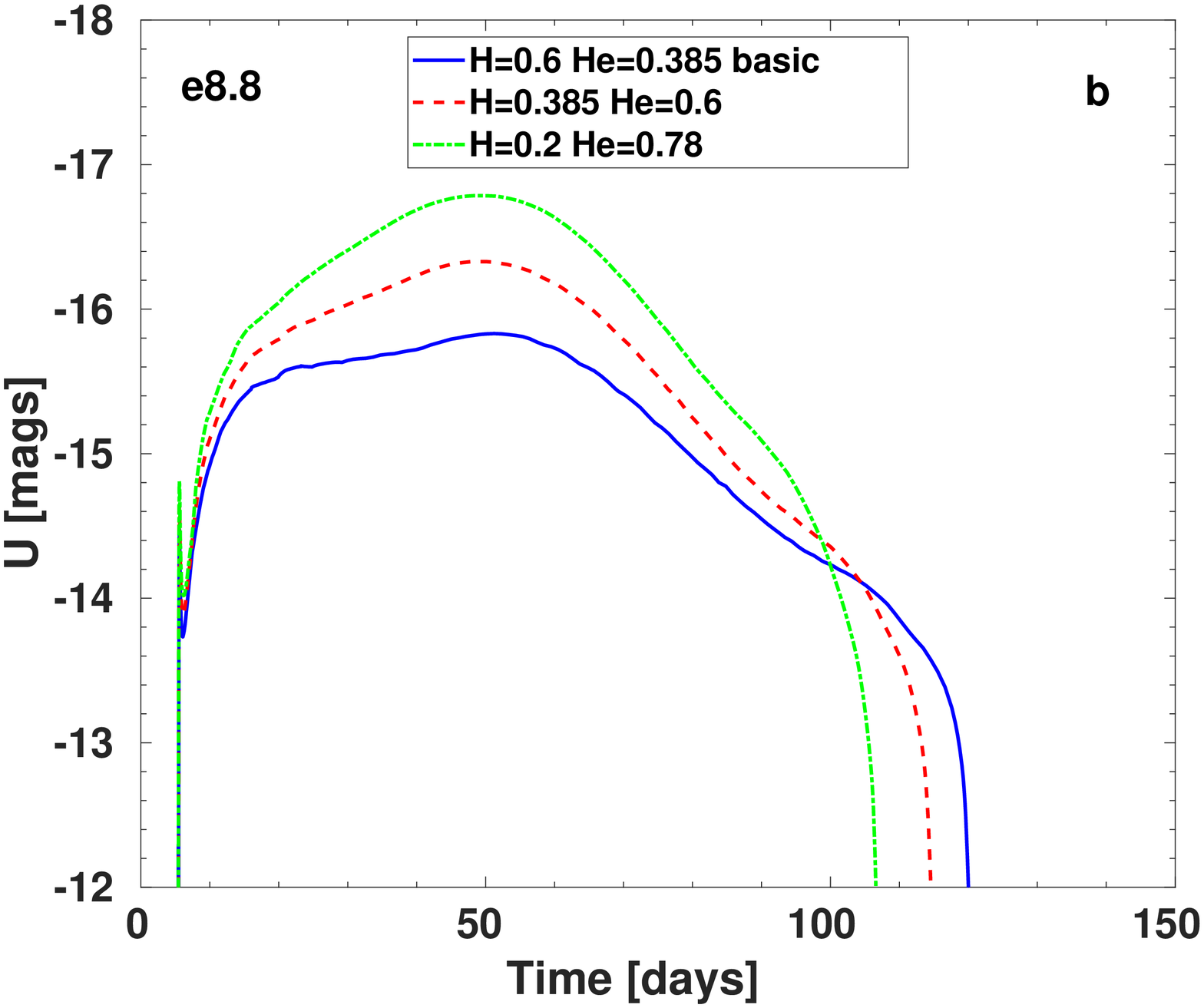}\\
\vspace{4mm}
\includegraphics[width=0.5\textwidth]{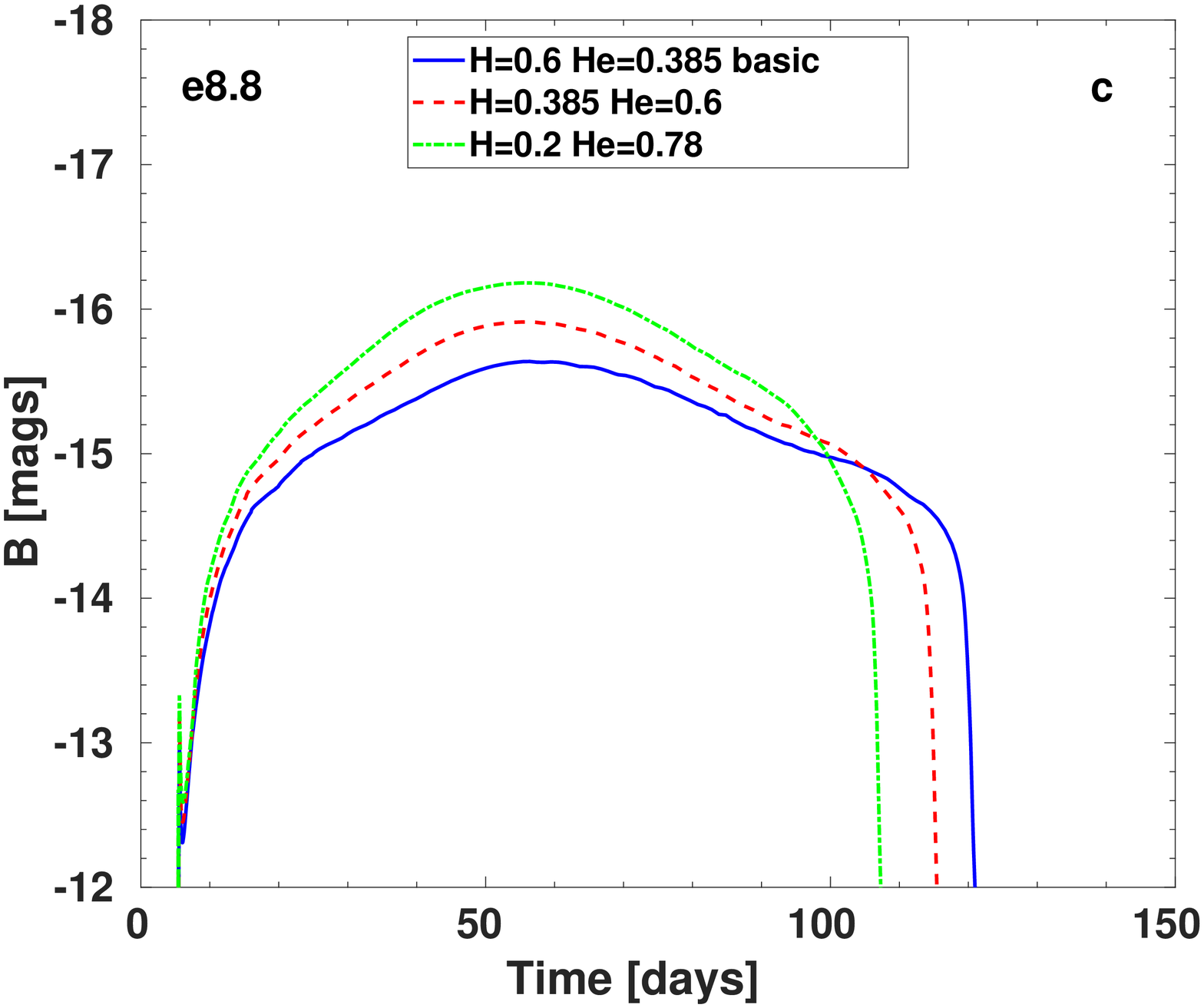}~
\hspace{1mm}\includegraphics[width=0.5\textwidth]{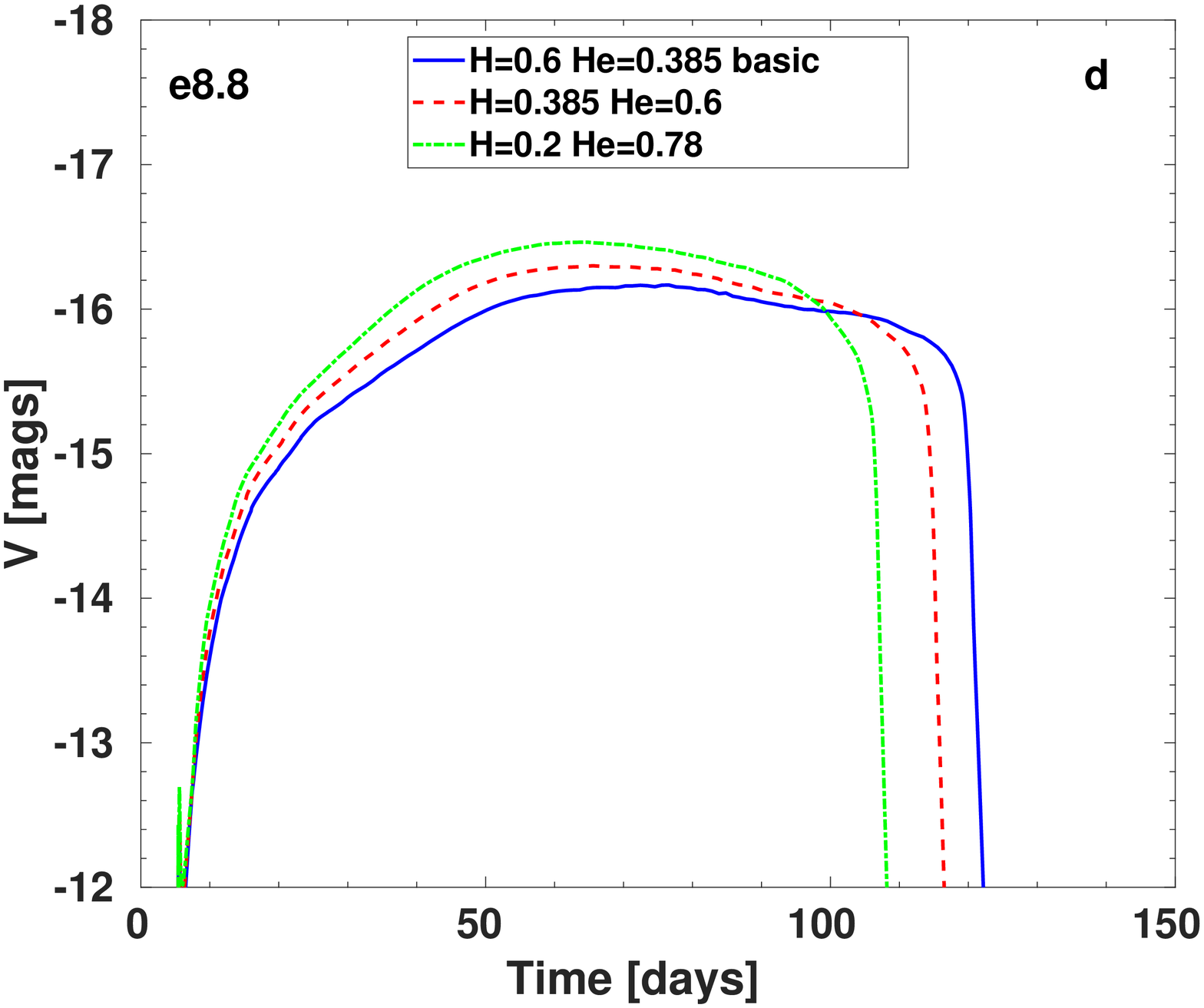}\\
\vspace{4mm}
\includegraphics[width=0.5\textwidth]{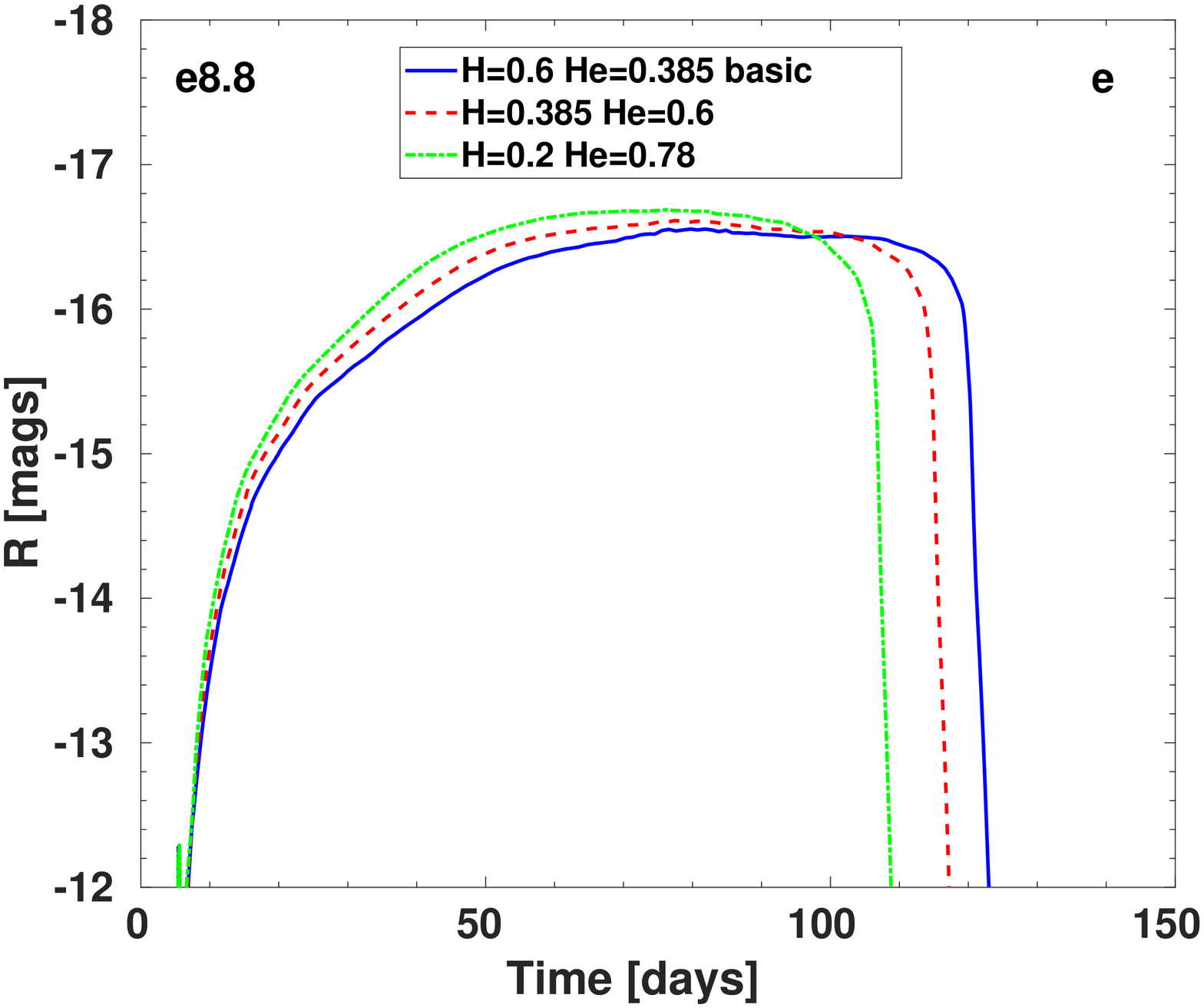}~
\hspace{3mm}\includegraphics[width=0.49\textwidth]{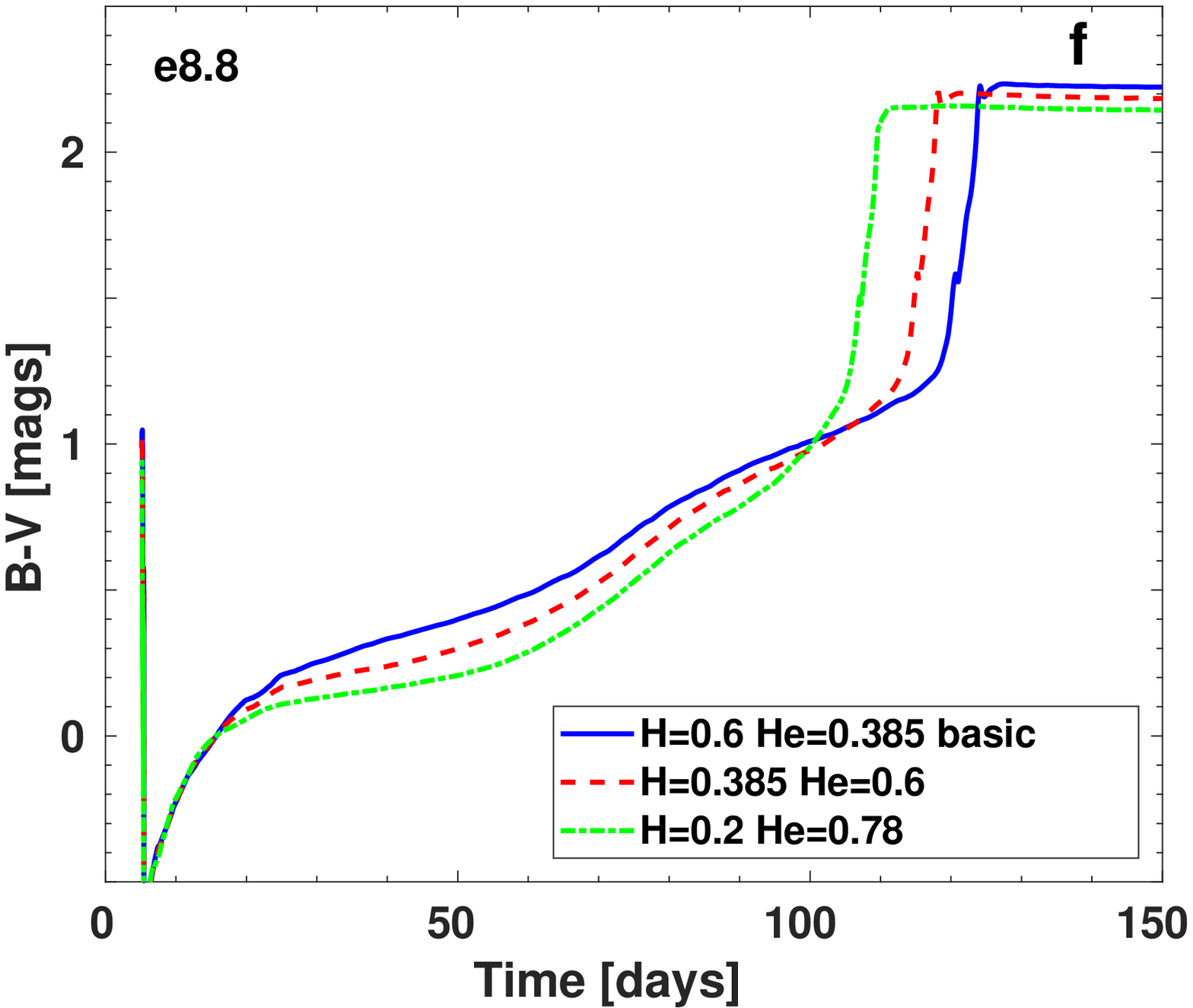}\\
\caption{Dependence on the hydrogen-to-helium-ratio: Bolometric and broad-band light curves, and the \emph{B}--\emph{V}
colour for the subset of runs based
on the model e8.8 with different hydrogen-to-helium-ratio: 0.385:0.6,
0.385:0.6, and 0.2:0.78.}
\label{figure:HHedepend}
\end{figure*}

Different works \citep[e.g., ][ and other studies]{1987ApJ...322..206N,1999ApJ...515..381R,2007A&A...476..893S} 
report that one of the distinct
properties of the evolution of stars in the narrow mass range around 8~\Msun{},
is a sequence of carbon-shell burning flashes which lead to 
the dredge-out episode. This later results in complete destruction of the helium
layer and macroscopic injection of helium into the hydrogen-rich envelope. The
stellar models consequently tend to have a decreased hydrogen-to-helium
ratio in the envelope (see, however, \citealt{2015ApJ...810...34W}). \citet{2013ApJ...771L..12T}
explored the consequence of different hydrogen-to-helium ratios on SN light
curves. In their study, the relative hydrogen abundance in the envelope was set to 0.7, 0.5, and 0.2. 
We carried out a similar study, this time assuming two different hydrogen abundances
in addition to the default value:
0.385 and 0.2 (the default value of the hydrogen abundance is 0.6). The resulting
light curves are shown in Figure~\ref{figure:HHedepend}. The impact of
a different hydrogen-to-helium ratio is the same as found by
\citet{2013ApJ...771L..12T} and similar to that found by
\citet{2009ApJ...703.2205K} for normal SNe~IIP. The lower hydrogen abundance leads to
a slightly shorter and brighter plateau. This is explained by the lower electron abundance,
in this case. The lower electron abundance leads to a lower electron-scattering opacity which governs the dynamics of the
receding photosphere. According to \citet{1982Natur.299..803N}, the hydrogen-to-helium ratio
in the Crab nebula, which is believed to originate from an ECSN explosion,
ranges between 0.125 and 0.625 with the hydrogen fraction in the range 0.2--0.3.
Therefore, the green light curves in Figure~\ref{figure:HHedepend} are more favourable for
this ECSN candidate. The \emph{B}--\emph{V} colour differs not significantly between
variants of e8.8 with different hydrogen-to-helium ratio, showing a scatter of 0.2~mag during the plateau phase.
Otherwise the behaviour of the light curves with different hydrogen-to-helium
ratios is similar.

\subsection[Dependence on the Tracer]{Dependence on the assumed yield of the Tracer material}
\label{subsect:tracer}

\begin{figure}
\centering
\includegraphics[width=0.5\textwidth]{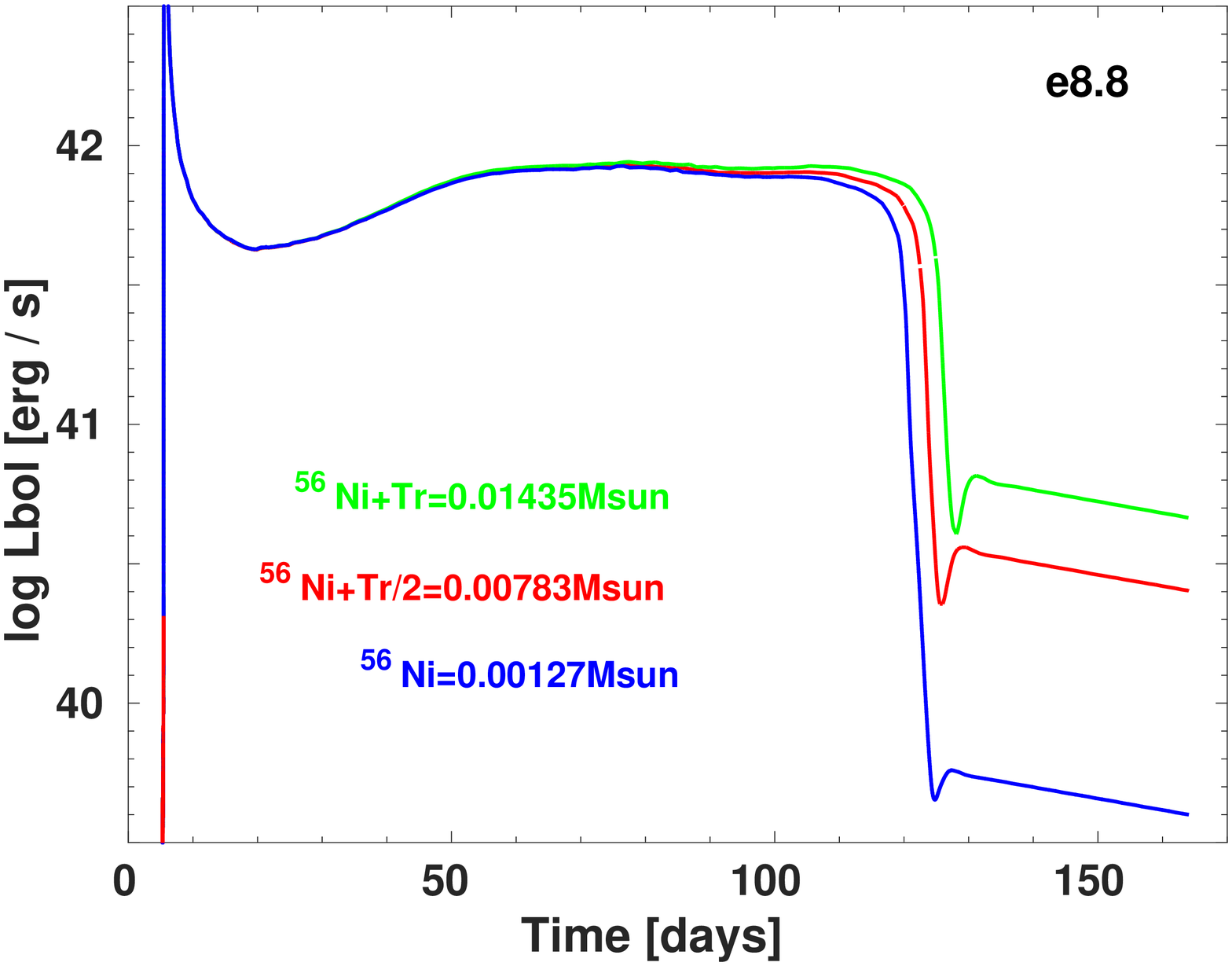}
\caption{Bolometric light curves for the subset of runs based on the model e8.8 with
different $^{56}$Ni{} mass: ``pure'' $^{56}$Ni{}, $^{56}$Ni{}+Tracer/2, and
$^{56}$Ni{}+Tracer, which correspond to 0.00127~\Msun{}, 0.00783~\Msun{},
and 0.01435~\Msun{}.}
\label{figure:Tracer}
\end{figure}

The core-collapse explosion simulations for our models were carried out with
the \verb|PROMETHEUS| code, which provides
a small 23-isotope (\verb|VERTEX-PROMETHEUS|) and a 16-isotope
(\verb|PROMETHEUS-HOTB|) nuclear network \citep[for details, see ][]{2020MNRAS.496.2039S}.
The production of iron-group elements in those studies is governed by the
reduced set of nuclear species used in the treatment of nuclear statistical
equilibrium and in a simplified nuclear alpha-reaction network. 
The latter provides an approximate estimate of the $^{56}$Ni{}
yield. However, the exact amount of $^{56}$Ni{} depends on the electron
fraction (or neutron-to-proton ratio) in the ejecta, which is not accurately
determined with the approximate neutrino transport used in the \verb|PROMETHEUS-HOTB| simulations
of the model e8.8. In neutron-rich conditions (electron fraction $Y_e<0.49$)
the code produces a so-called ``tracer'' nucleon that traces neutron-rich
nuclei. With more accurate $Y_e$, some fraction of this tracer could
actually be radioactive nickel $^{56}$Ni{}. To account for this uncertainty,
we run two additional simulations
for the same hydrodynamical profile of the default model e8.8 for which we
set the $^{56}$Ni{} yield to be the following: $^{56}\mathrm{Ni}+\mathrm{Tracer}/2$
and $^{56}\mathrm{Ni}+\mathrm{Tracer}$. We show the result of the simulations
in Figure~\ref{figure:Tracer}. Different total amounts of
radioactive nickel $^{56}$Ni{} with the same shape of distribution throughout the
ejecta largely affect the luminosity of the radioactive tail and the
extension of the plateau \citep{2019MNRAS.483.1211K}. Inclusion of the
entire mass of the Tracer into the $^{56}$Ni{} mass leads to higher mass of
the radioactive material, i.e. reduces the drop between the plateau and the
tail, and makes the ECSN bolometric light curve more similar to
that of normal SN~IIP of a massive progenitor (10\,--\,20~\Msun{}).
However, the overall spectral evolution and colours remain the same as for
the default model e8.8.

\subsection[Dependence on the progenitor radius]{Dependence on the progenitor
radius. The case of e8.8}
\label{subsect:radius}

\begin{figure*}
\centering
\hspace{9mm}\includegraphics[width=0.49\textwidth]{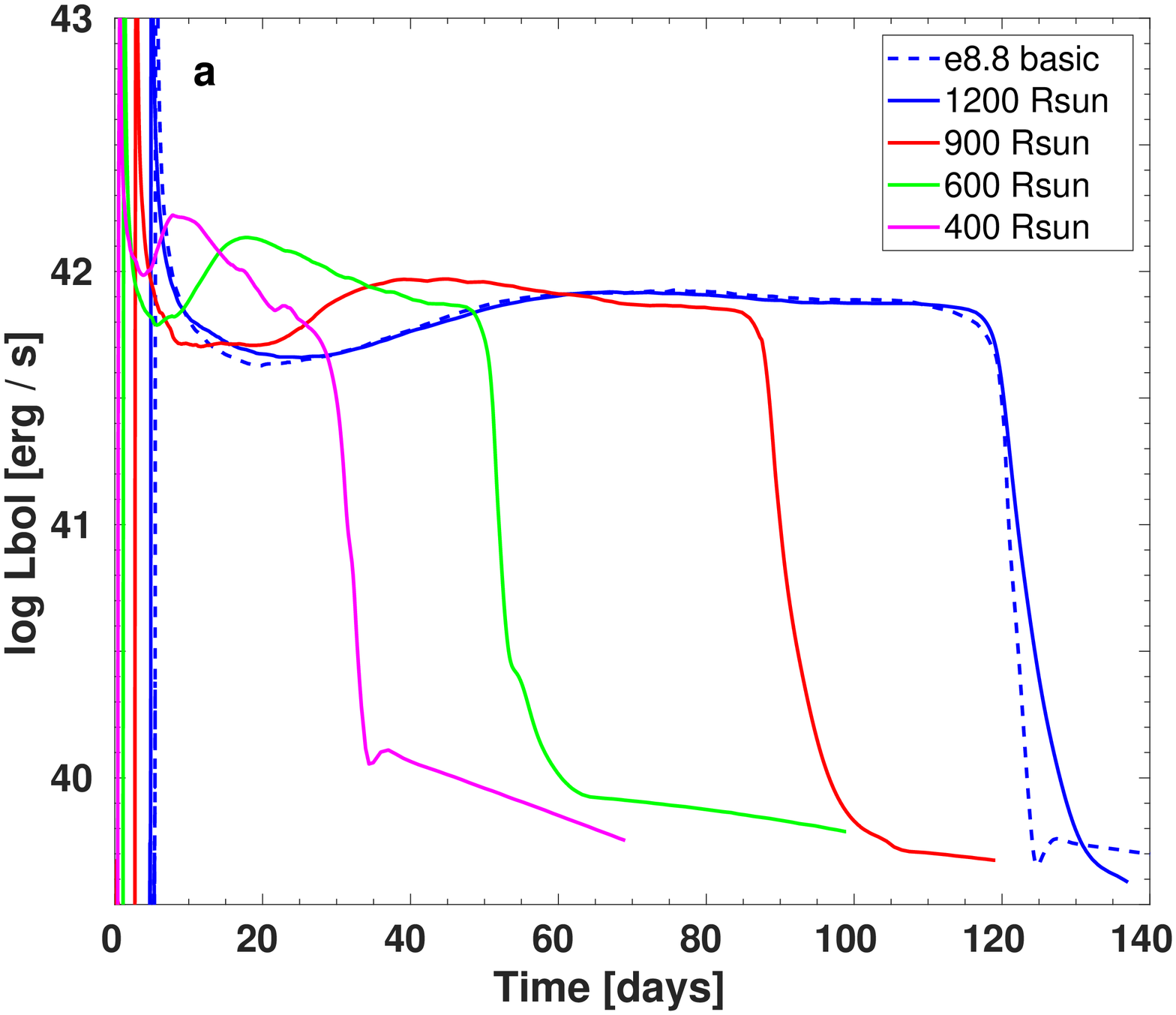}~
\hspace{3mm}
\includegraphics[width=0.5\textwidth]{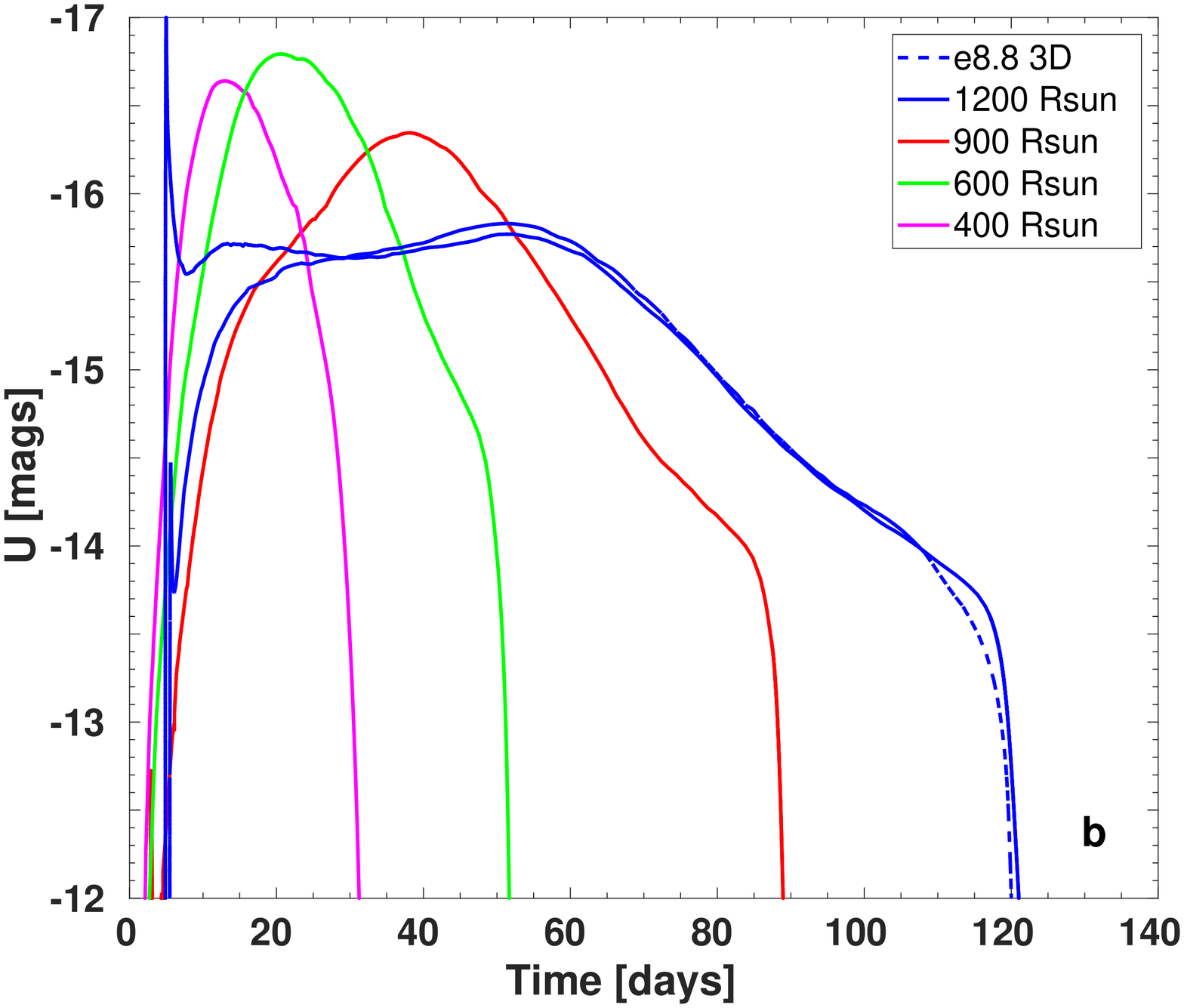}\\
\vspace{5mm}
\includegraphics[width=0.5\textwidth]{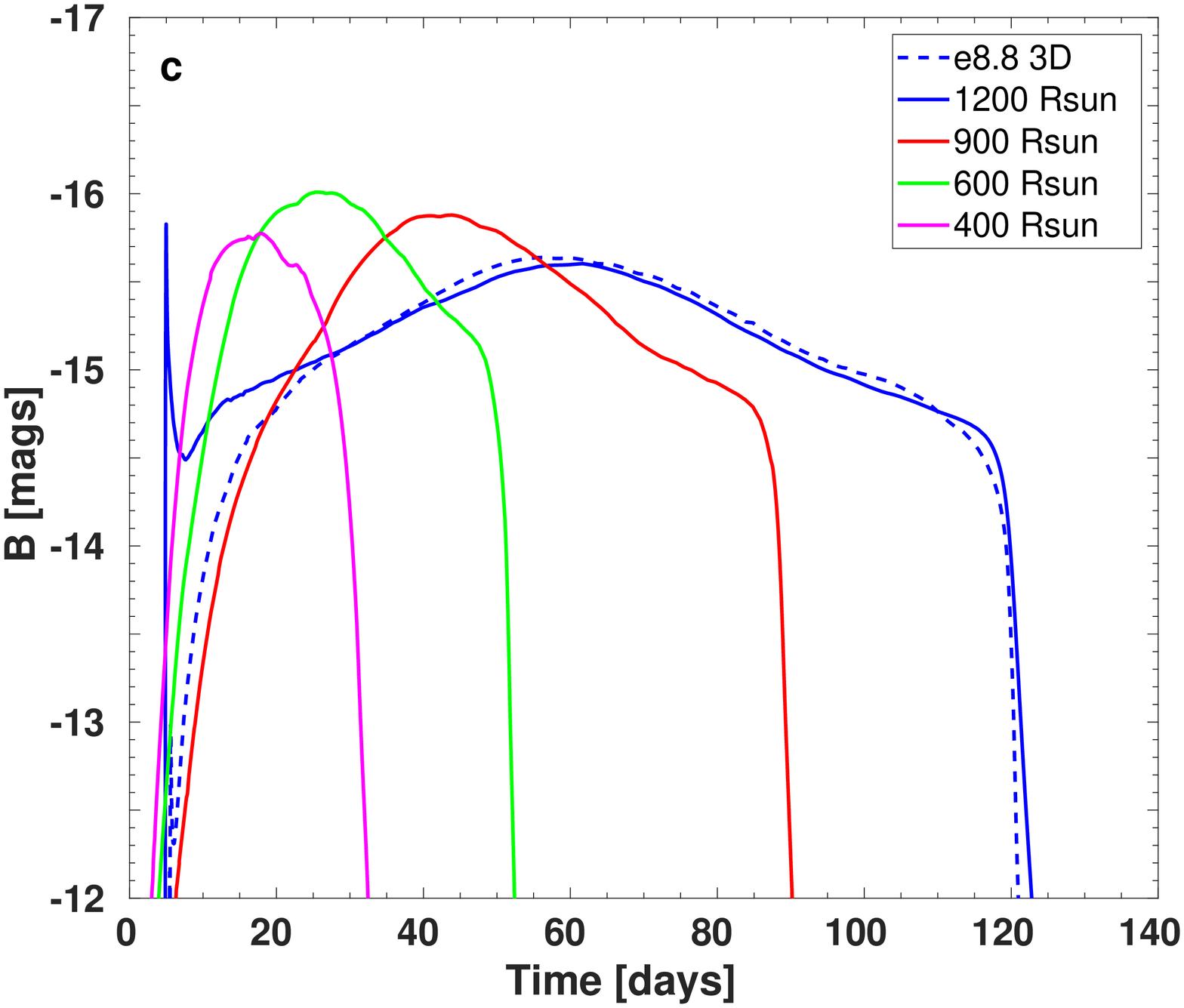}~
\hspace{2mm}
\includegraphics[width=0.5\textwidth]{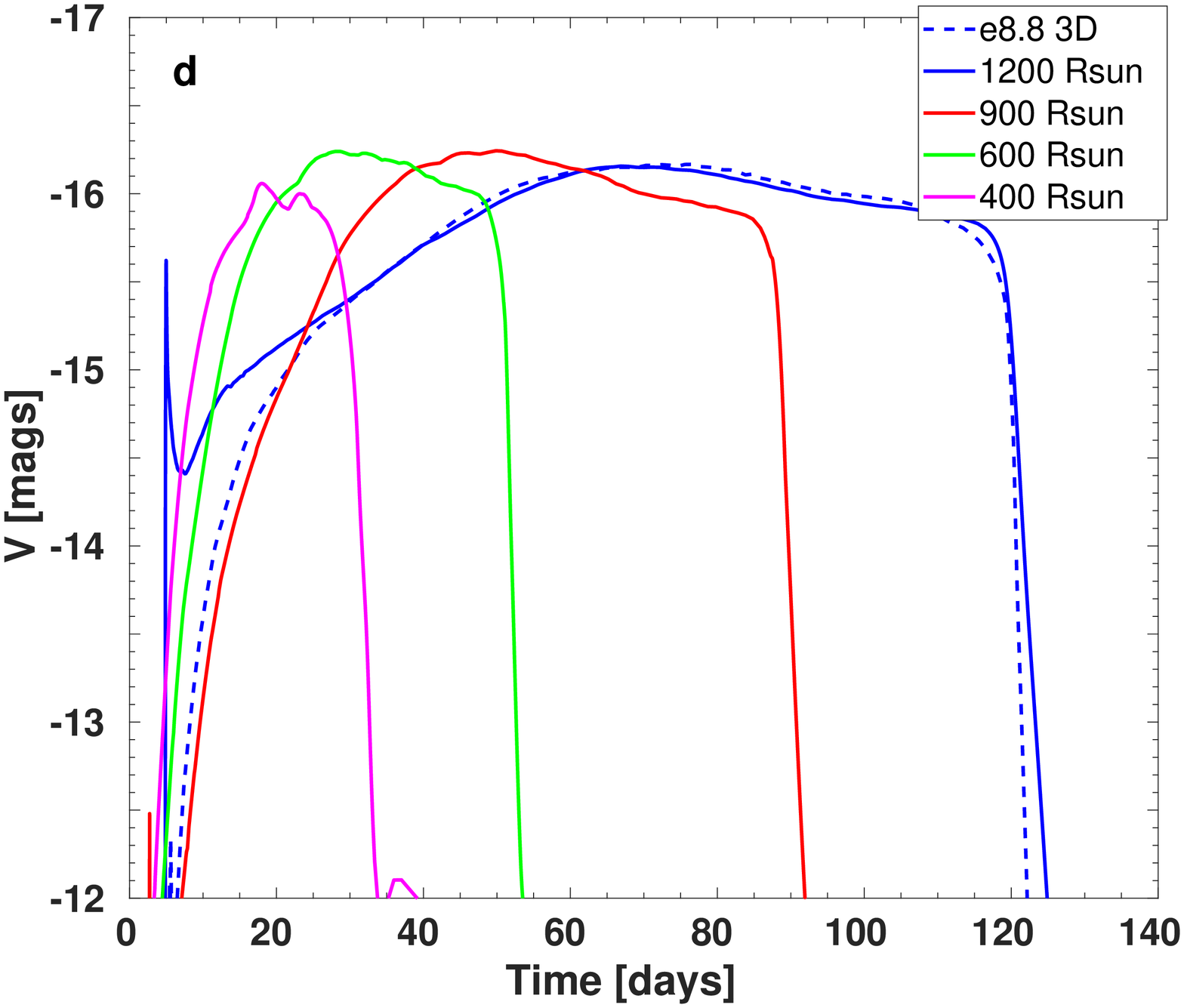}\\
\vspace{4mm}
\includegraphics[width=0.5\textwidth]{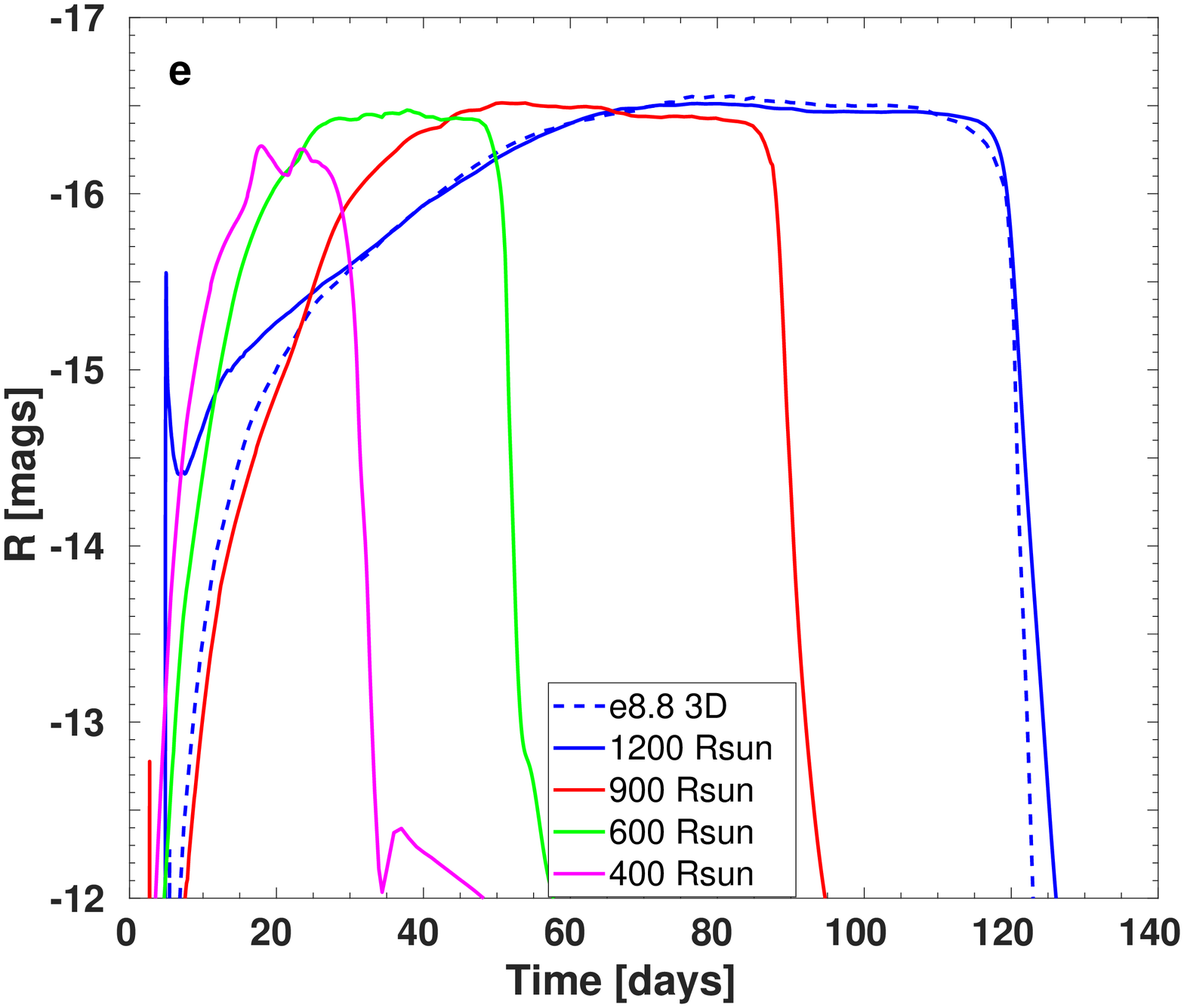}~
\hspace{4mm}
\includegraphics[width=0.495\textwidth]{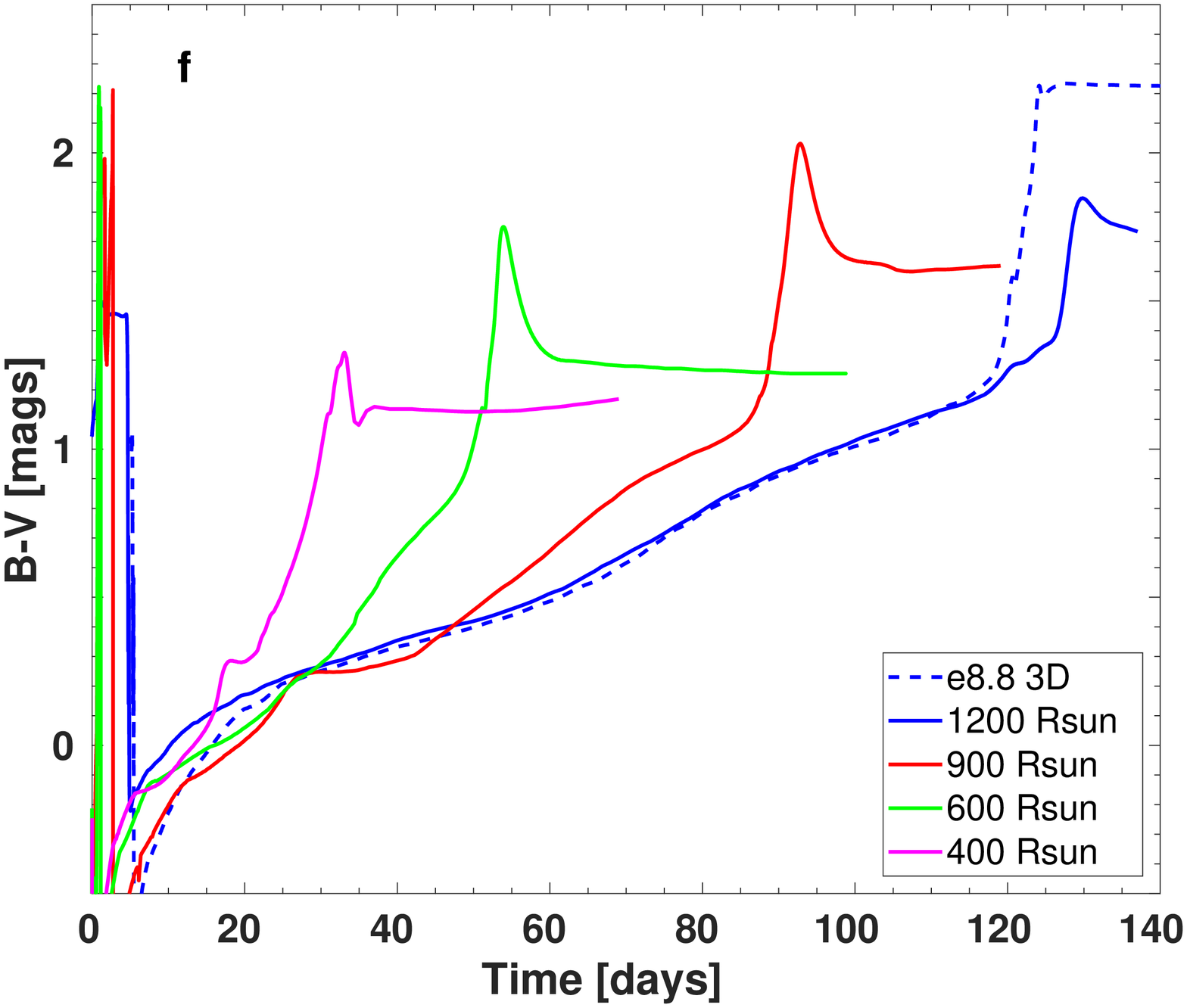}\\
\caption{Bolometric light curves for the subset of runs based on the stellar
evolution model e8.8 with different degree of truncated envelope.}
\label{figure:radius}
\end{figure*}

It is known that the large fraction of stars are born in binary, triple or multiple
systems \citep{1992ApJ...391..246P}. Similar to the famous Algol paradox,
stars lose its mass via critical Roche surface \citep{1998A&AT...15..357P}.
The stellar evolution of individual stars is strongly affected if they are
part of a close binary. Therefore,
it is adequate to admit some degree of uncertainty to the
mass-loss. Additional mass-loss happens via mass transfer
in a binary system, which is the channel for some initially hydrogen-rich
low-metallicity massive stars to result in
hydrogen-free supernovae SN~Ib/c \citep{2012A&A...544L..11Y}.
\citet{2012A&A...538L...8G} show that increasing the wind mass-loss rate by
a factor of 3 to 10 shrinks the progenitor, i.e. makes the more compact star.
Multipling the rate of wind mass-loss may mimic the binarity, i.e. the
enhanced mass-loss happening in a binary via Langrangian point L1.
Specifically, ECSNe may occur in binaries
\citep{2008MNRAS.384.1109E,2017ApJ...850..197P,2018A&A...614A..99S}.
\citet{2015MNRAS.451.2123T} and \citet{2019A&A...622A..74J} show that 
a star that would otherwise result in an ECSN will instead result in an
ultra-stripped SN if part of a close binary.
Losing mass via close binary interaction, i.e. via the Roche lobe overflow, 
a star becomes more compact. 
Hence, we did additional subset of runs based on the stellar evolution model
e8.8 which has radius of 1200~\Rsun{} at the moment of collapse. Three
another truncated models have radii of 900~\Rsun{}, 600~\Rsun{}, and
400~Rsun{}. We note that the main model e8.8 discussed above was evolved
with \verb|PROMETHEUS| up to the moment of shock breakout and then was
mapped into \verb|STELLA|. Truncating
the main profile (the \verb|PROMETHEUS| output) leads to cut in the total energy budget. Therefore, we used the
stellar evolution 1D output which was mapped into \verb|PROMETHEUS| prior to the
collapse. We detonate it with the thermal bomb method with according
explosion energy. We note that the amount of explosion energy was set
in a way to allow the subset models to reach the same terminal kinetic
energy of 0.86$\times\,10^{\,50}$~erg which is the explosion energy of the main
model e8.8 in the study. The reduction of the radius by cutting numerically
leads to some degree of inconsistency, since the star is supposed to relax into a
new thermodynamical equilibrium. Another unavoidable side-effect consists in
reduction of the ejecta mass. Hence the total mass for the
truncated sub-models are: 1.8~\Msun{}, 2.4~\Msun{}, 4~\Msun{} for
400~\Rsun{}, 600~\Rsun{}, and 900~\Rsun{} cases, respectively.

In Figure~\ref{figure:radius}, the resulting bolometric, broad-band light curves
and the \emph{B}-\emph{V} colour are
shown. The artificially detonated stellar evolution model (labelled ``1200~Rsun'') displays the same
curve as the main model e8.8 exploded in a self-consistent manner (blue
solid and dashed curves). There is some difference between these
curves due to a few reasons. First, the detonated evolutionary model does not have any applied
macroscopic mixing of chemicals. Second, \verb|PROMETHEUS| simulations
were performed without taking radiation transport into account while
\verb|STELLA| does hydrodynamics coupled with radiation transport.
I.e. this leads to a difference in propagation of the radiation-dominated
shock. Third, \verb|PROMETHEUS| does not include heating from radioactive
{}$^{56}$Ni, while \verb|STELLA| does. 
The presence of {}$^{56}$Ni results in variation in velocity, temperature and density
field which is called ``Ni-bubble'' effect and leads to variation in, e.g., velocity
upto 10\,--\,15\,\% in the region where {}$^{56}$Ni mass fraction is close to unity
\citep{2017MNRAS.464.2854K}. Fourth, the \verb|PROMETHEUS| 3D output was angle-averaged to
be suitable for mapping into \verb|STELLA| which leads to the slight variation
in hydrodynamical profiles and may cause some diversity in the resulting
radiation transfer simulations. 

As seen in Figure~\ref{figure:radius}, the more compact the progenitor the shorter the
plateau, which is consistent with Equation~\ref{equation:Popov} \citep{1993ApJ...414..712P} and
numerical experiments by \citet{2004ApJ...617.1233Y}. However,
the plateau duration has the major impact from the ejecta mass which is
connected to the reduction of the radius.
The model e8.8 being truncated to 400~\Rsun{} has a very short living light
curve lasting only 30~days, although it still retains hydrogen-rich envelope
with the total mass of hydrogen of 0.3~\Msun{}. 
The colour of the more truncated models tends to be bluer since 
the hotter region of the ejecta gets closer to the outer edge. 
The higher temperature also explains the bump in the bolometric light curves
for the cases of 400~\Rsun{} and 600~\Rsun{}. For the same
reason \emph{U}-band light curves for the 400~\Rsun{} and 600~\Rsun{} cases
are relatively luminous.
We conclude that ECSNe in the binaries have fastly declining low-luminosity light curves
and will be easily missed even being exploded in the close vicinity.
This is the reason that there is no solid detection of an ECSN yet.


\section[Comparison to a normal SN~IIP model]{Comparison to a normal SN~IIP model}
\label{sect:comp}

\begin{figure*}
\centering
\includegraphics[width=0.5\textwidth]{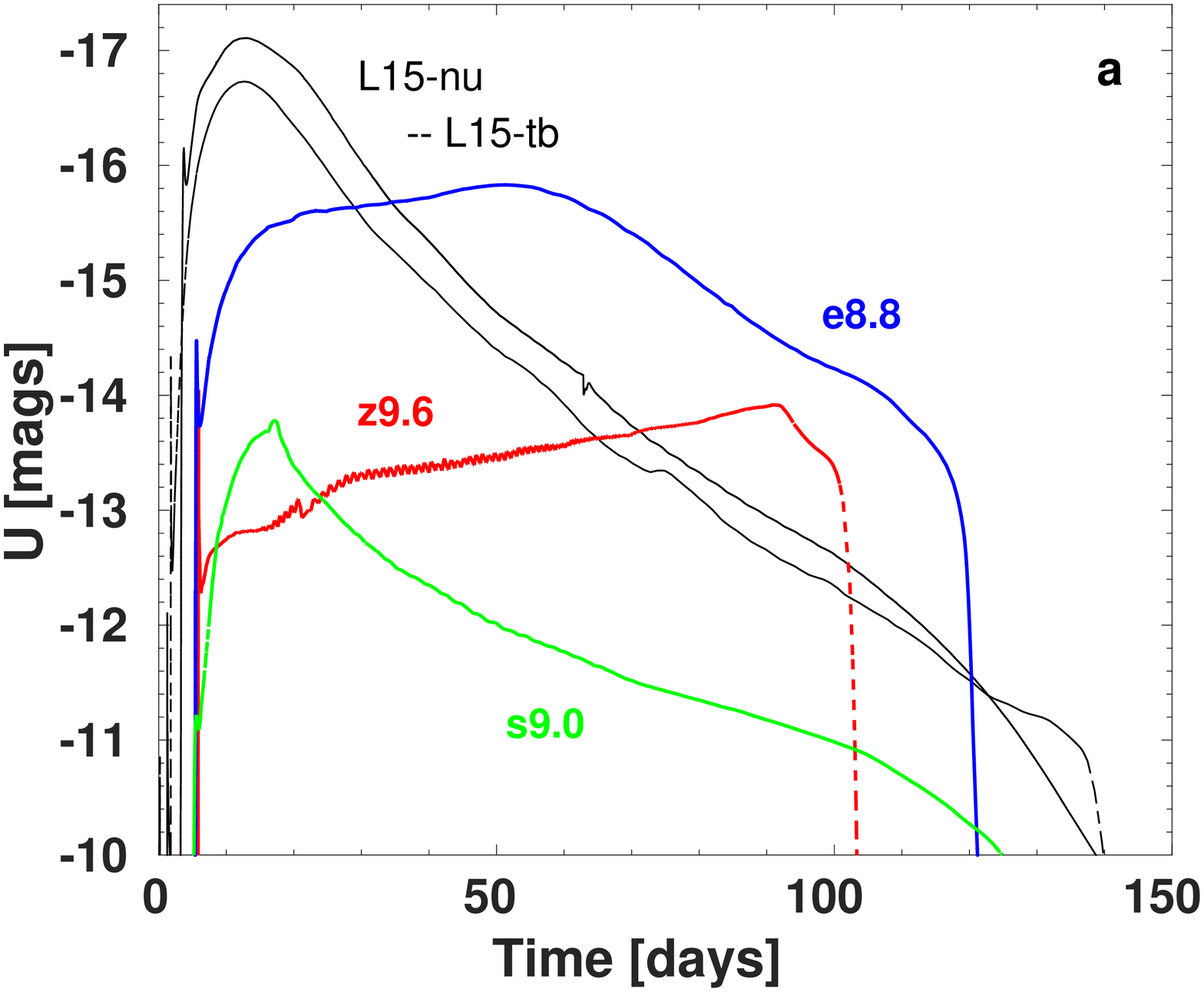}~
\includegraphics[width=0.5\textwidth]{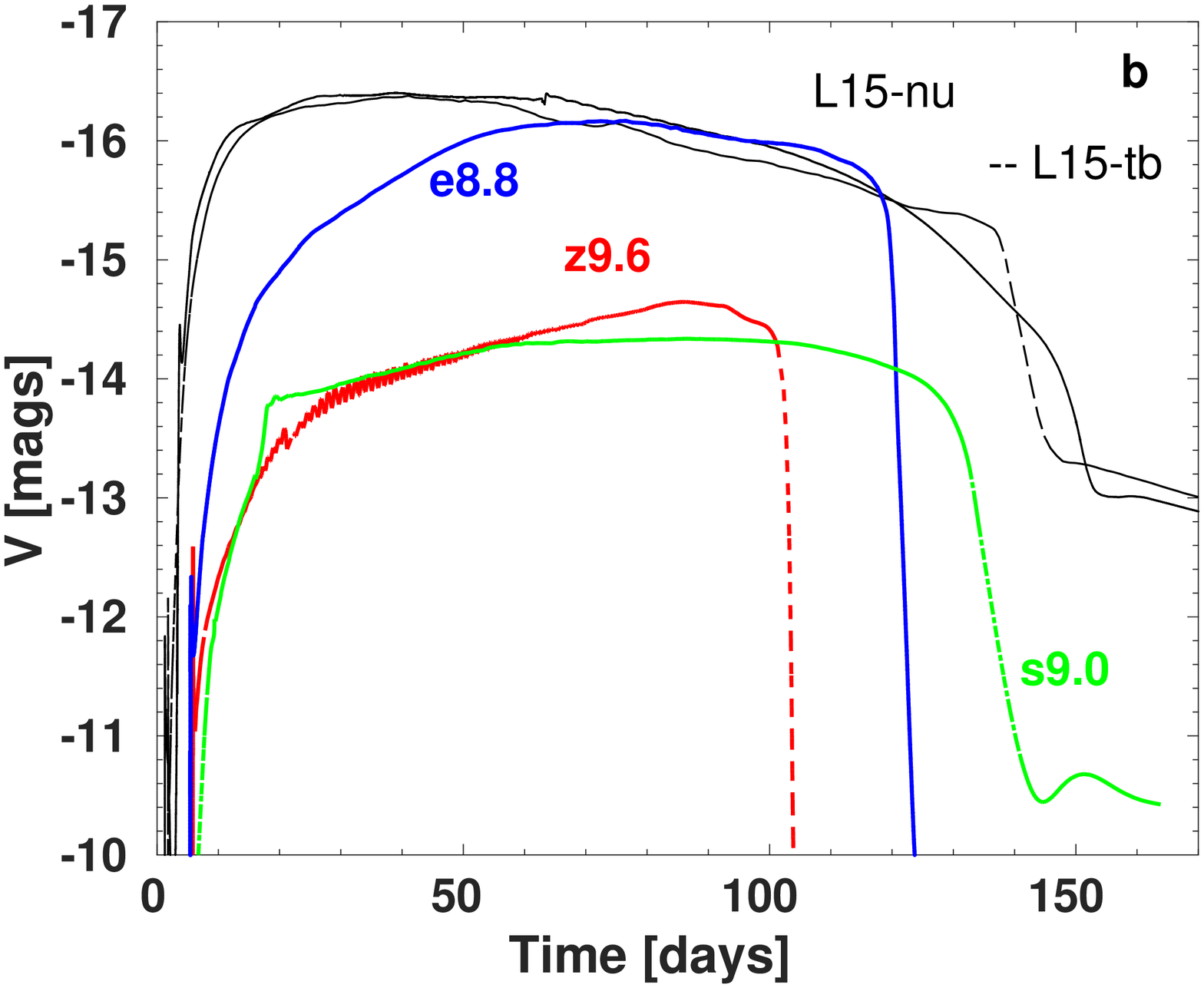}\\
\caption{\emph{U} and \emph{V} broad-band light curves for the models e8.8, z9.6, 
s9.0, and L15-nu/L15-tb.}
\label{figure:bands}
\end{figure*}

\emph{U} and \emph{V} broad-band light curves for three default models
from our study and the reference CCSN model L15-nu/L15-tb are shown in
Figure~\ref{figure:bands}. The \emph{U} magnitude of the model z9.6 evolves
significantly differently to other models due to its lack of metals. We
study the effect of metallicity and discuss our results in Section~\ref{subsect:Zdepend}
above.  The light curves of L15-tb and L15-nu in \emph{U}-band decline 2~mags during
50~days, i.e. evolves typically for the canonical SNe~IIP. 
s9.0 \emph{U}-band light curve behaves the same way as the reference
CCSN model L15-nu/L15-tb with a systematically lower luminosity overall.
There is no appreciable difference between plateau-like light curves in \emph{V} band for
the ECSN model e8.8 and the low-mass CCSN models z9.6 and s9.0 in comparison to the
reference CCSN model L15-nu/L15-tb.

The colours \emph{B}--\emph{V}, \emph{V}--\emph{R}, and \emph{V}--\emph{I}
are bluer for z9.6 and e8.8 compared to the models s9.0 and L15-nu/L15-tb.
This is a conquence of the lack of metals in the model z9.6 and the unusual colour temperature
evolution for the model e8.8. We discuss the latter below.

\begin{figure}
\centering
\includegraphics[width=0.5\textwidth]{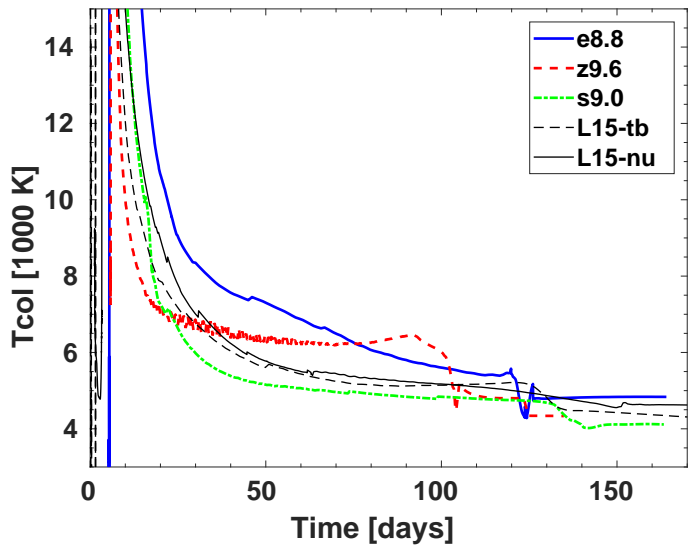}
\caption{Colour temperature evolution for the models e8.8, z9.6, 
s9.0, and L15-nu/L15-tb.} 
\label{figure:Tcol}
\end{figure}

The colour temperature evolution for all models is shown in
Figure~\ref{figure:Tcol}. The colour temperature is the black body
temperature as estimated from the least-square method. It serves
as the indicator of the maximum in the spectral energy distribution, i.e.
represent the frequency where the major flux is radiated.
As was discussed above, recombination
sets in at relatively late times in the model e8.8 compared to the other
models. This is explained by a strong dependence of the
``recombination time'' on the radius of the progenitor and energy of the
explosion (see Equation~\ref{equation:Trecomb}). 
The distinguishing property of ECSN progenitors is the large
radius, and generally ECSN explosions are low energetic, therefore, the
recombination time tends to be relatively long. Specifically the radius of model e8.8 
is 1200~\Rsun{}, quite large even for an average red
supergiant. The explosion energy of the reference explosion
model e8.8 is 0.1~foe. According to Equation~\ref{equation:Trecomb}
\citep{Shussman2016,2020MNRAS.494.3927K},
recombination is established at day~66 for the model e8.8, day~22 for the
models z9.6 and L15, and at day~38 for the model s9.0. 

Hence the main feature which distinguishes ECSN explosions from CCSN
explosions are the behaviour of the blue flux, i.e. \emph{U} and \emph{B}
magnitude, colour temperature, and colours. The specific 
features of e8.8's observables are:
\begin{enumerate}
\item the light curve in the \emph{U} and \emph{B} bands rising during the first
50~days and then slowly declining, 
\item the light curve in \emph{V} and other redder bands
rising during the same phase of 50~days and then holding at a plateau before
the sharp drop to the low-luminosity tail powered by a small amount of the
radioactive nickel $^{56}$Ni{}.
\end{enumerate}
We emphasise that this unique behaviour is the
consequence of the relatively extended hydrogen-rich envelope and the absence of
helium and oxygen shells which in turn is the result of the unique stellar
evolution path of the super-AGB stars \citep[see e.g., ][]{2015MNRAS.446.2599D}.

The main differences between low- and normal-mass CCSNe are 
(1) an overall lower luminosity at the plateau, 0.5\,--\,1~dex for bolometric luminosity
or 1.5--2~mags in \emph{V} broad band, 
(2) a low-luminosity radioactive tail, and
(3) a large drop in luminosity between the plateau and the tail.
All of these features
are explained by the relatively lower explosion energy of low-mass explosions (below
0.1~foe) and a lower yield of radioactive nickel.

\begin{figure*}
\centering
\includegraphics[width=0.31\textwidth]{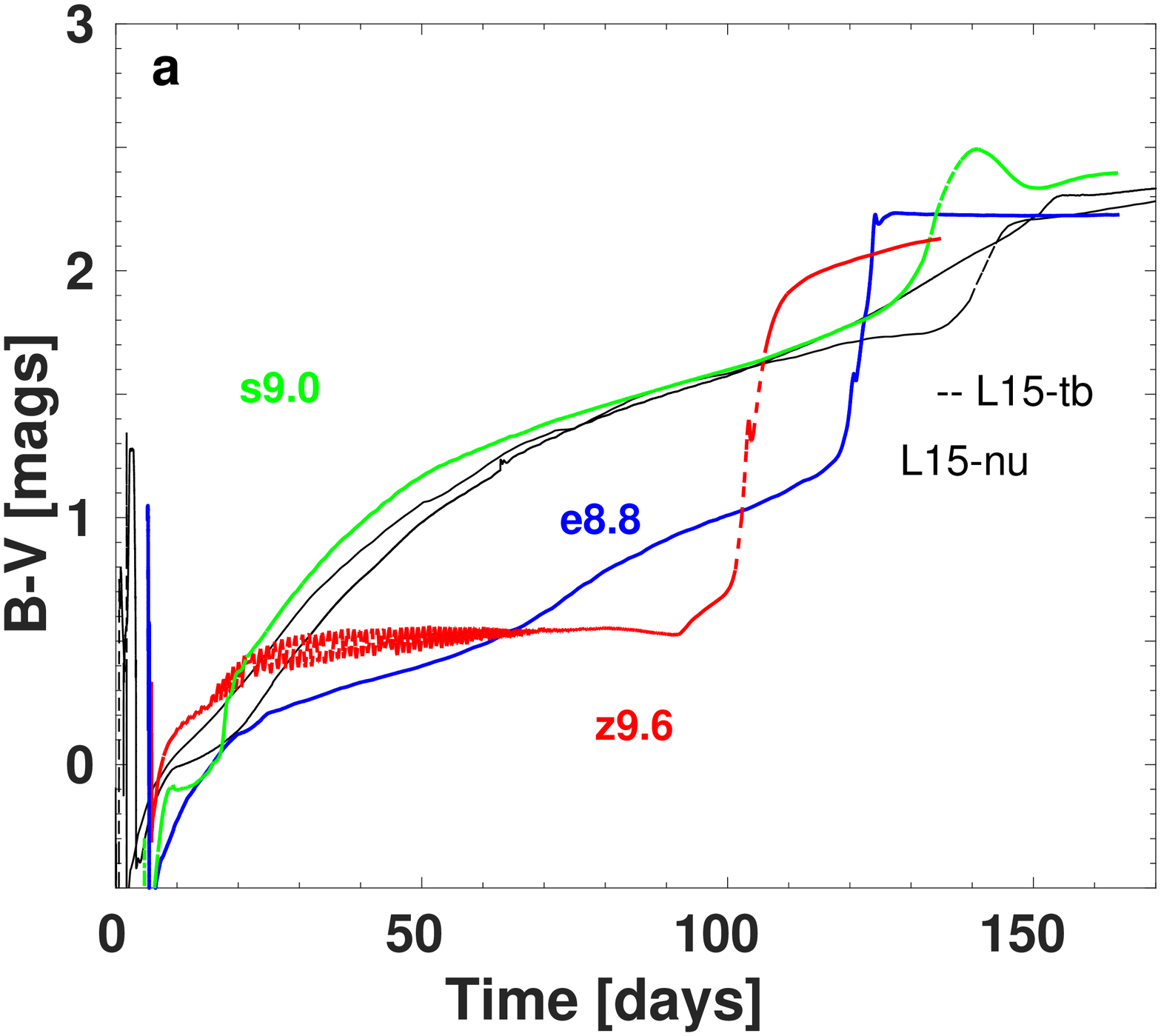}\hspace{3mm}
\includegraphics[width=0.31\textwidth]{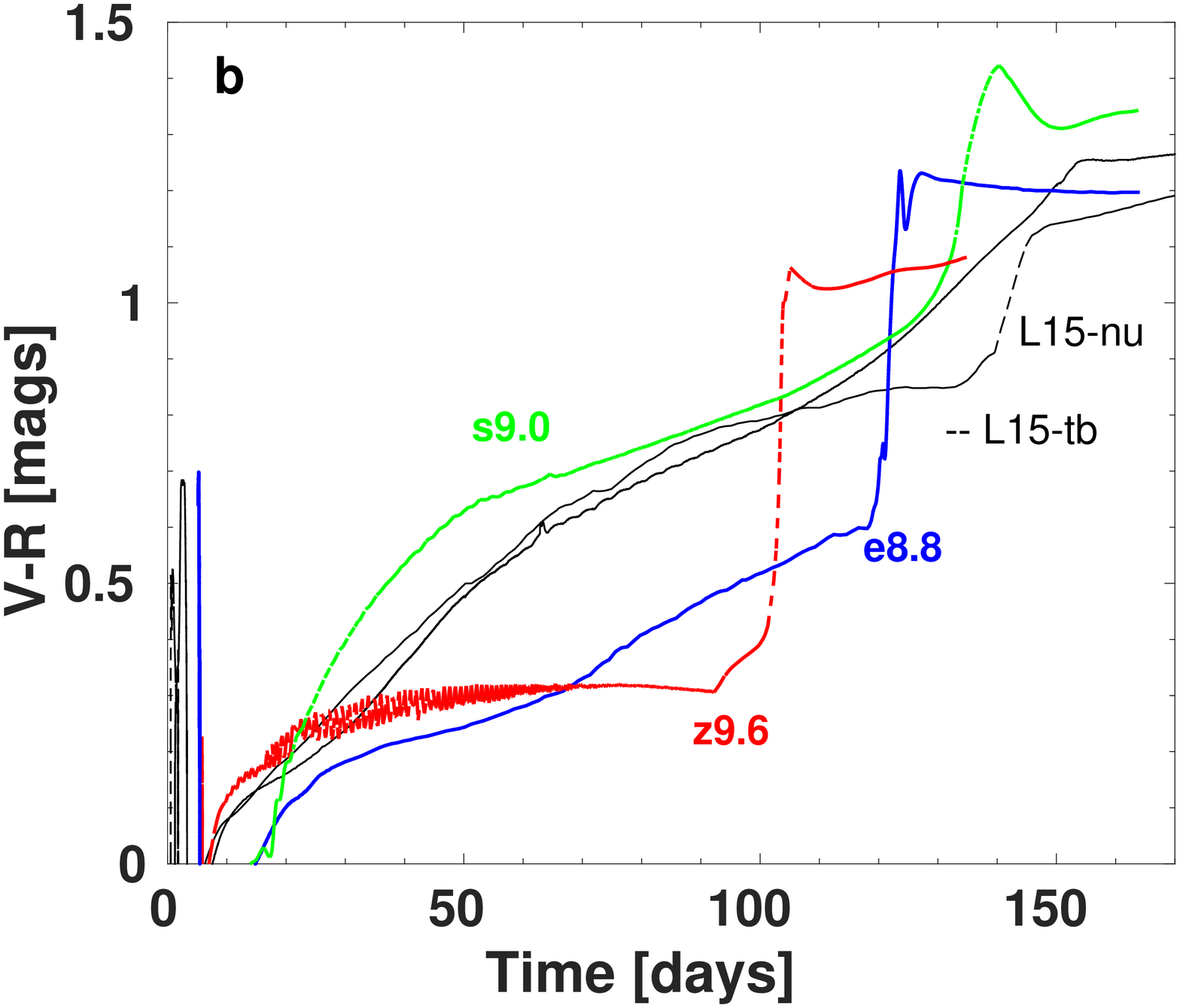}\hspace{3mm}
\includegraphics[width=0.31\textwidth]{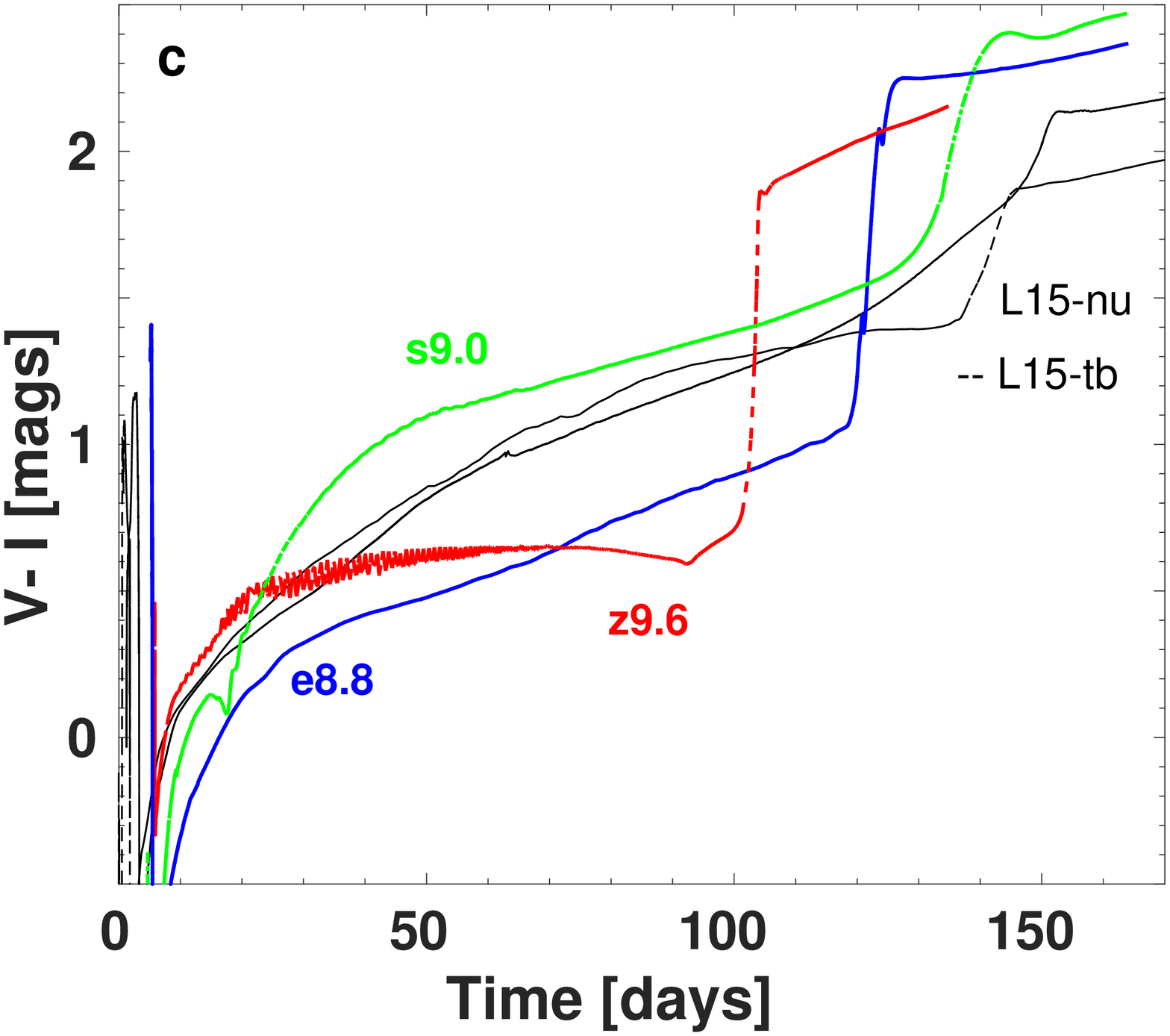}\\
\caption{\emph{B}--\emph{V}, \emph{V}--\emph{R}, and \emph{V}--\emph{I} 
colours for the models e8.8, z9.6, s9.0, and L15-nu/L15-tb.}
\label{figure:colour}
\end{figure*}


\section[Application to observations]{Application to observations}
\label{sect:observe}

In this section we consider the results of our radiative transfer
simulations for the ECSN model e8.8 and the low-mass CCSN models z9.6 and s9.0
in the context of observations. Particularly, we aim to find possible
candidates matching our calculations. Further, we aim to point out the criteria for
distingushing properties,
specifically for the ECSNe and low-mass CCSNe. Immediately we encounter an obvious
problem: namely that the number of detected and followed-up SNe increases every year at an
exponential rate \citep{2013PASP..125..749G}. Therefore, in our analysis we
opt to compare our synthetic
light curves to a representative sample and draw some general conclusions.

\subsection[SN~2005cs]{SN~2005cs}
\label{subsect:2005cs}

\begin{figure}
\centering
\includegraphics[width=0.5\textwidth]{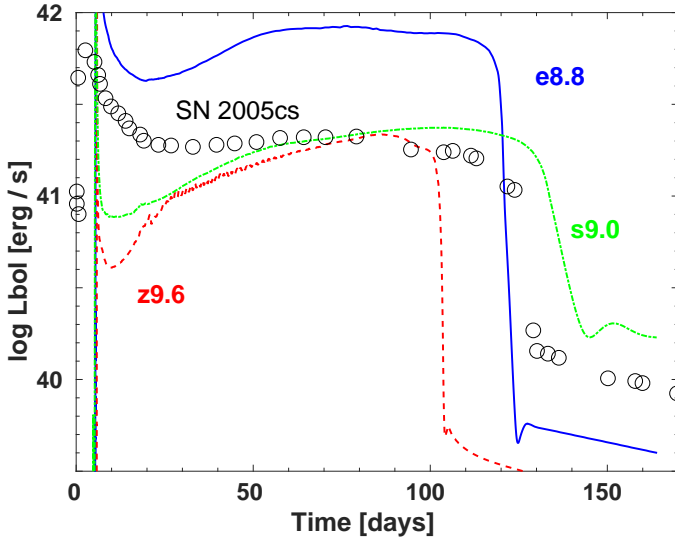}
\caption{Bolometric light curves for the models e8.8, z9.6 and s9.0 and
observed SN~2005cs \citep{2009MNRAS.394.2266P}.}
\label{figure:lbol2005cs}
\end{figure}

For example, low-mass CCSNe are expected to be low energy and to produce
a small amount of $^{56}$Ni leading to a low-luminosity plateau, and a large
drop towards the low-luminosity radioactive tail. Therefore, we choose
SN~2005cs as the first example for our comparison. This event is famous for its low plateau
luminosity and steep and pronounced drop towards the tail. We show the bolometric light curve
of SN~2005cs \citep{2009MNRAS.394.2266P} superposed on our simulated bolometric light curves
in Figure~\ref{figure:lbol2005cs}. It is apparent that none of the
models match the bolometric light curve of SN~2005cs. However, there are
some conclusion to be drawn on the general energetics and bolometric
parameters of this particular SN and its possible progenitor. (1) SN~2005cs is
likely not a ECSN explosion with an envelope and energy as model e8.8,
because the luminosity of the SN throughout the plateau phase
is about 0.2--0.5~dex lower than the luminosity of the ECSN model e8.8. This means
that the progenitor of SN~2005cs was likely more compact than our model
(with a radius of 1200~\Rsun{}), which is in agreement with the conclusion
drawn by \citet{2007MNRAS.376L..52E}. (2) The model s9.0 is likely more similar
to the real progenitor
of SN~2005cs. The remaining differences are explained either by the radius of the progenitor 
larger than the radius of the model s9.0 (408~\Rsun{}) or/and the explosion energy
being higher than $0.7\times10^{\,50}$~erg. The total mass of the radioactive 
$^{56}$Ni{} in SN~2005cs is likely lower than the 
0.005~\Msun{} contained in the model s9.0. \citet{2009MNRAS.394.2266P} proposed the following
parameters for the progenitor based on their simulations: radius
100~\Rsun{}, ejecta mass 11.1~\Msun{}, and explosion energy
$3\times10^{\,50}$~erg. Later \citet{2014MNRAS.439.2873S} revised these numbers
providing new parameters for SN~2005cs: 350~\Rsun{}, 9.5~\Msun{}, and 1.6$\times10^{\,50}$~erg. 
The photospheric velocity of SN~2005cs is at the same level
as in our models (Figure~\ref{figure:uph}), i.e. about
1000--2000~km\,s$^{\,-1}$. Hence, judging from the bolometric properties of
the SNe, the
progenitor is likely a low-mass star of moderate radius which
exploded at relatively low energy, but still higher than the low energy explosions in our study.

\begin{figure}
\centering
\vspace{1mm}
\hspace{-5mm}\includegraphics[width=0.5\textwidth]{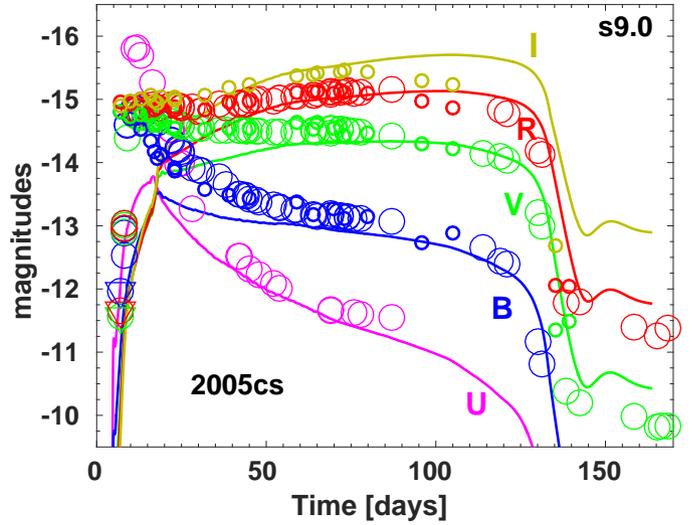}
\caption{Light curves for the model s9.0 and
observed SN~2005cs in broad bands 
Small and large circles correspond to the observed data taken from
\citet{2006A&A...460..769T} and \citet{2009MNRAS.394.2266P}, respectively.
Triangles present the upper limits for observed magnitudes.}
\label{figure:bands2005cs}
\end{figure}

We show the broad-band magnitudes of the
model s9.0 and SN~2005cs in Figure~\ref{figure:bands2005cs}. 
As discussed in the previous sections, \emph{U}
magnitude is a very unique indicator of many model parameters, such as iron and
nickel content in the ejecta and the radius of the progenitor. From the plot,
we conclude that the model s9.0, which is the most similar to a normal SN~IIP
judging from its \emph{U} band light curve, still evolves too shallowly. On the
one hand, this
is explained by the fact that radioactive nickel is mixed extensively in the
ejecta of s9.0, retaining flux in the blue bands to later times. On the other hand,
increased short wavelength flux
might be explained by a low stable iron content in the hydrogen-rich
envelope. However, as the model s9.0 is computed with solar metallicity,
and the contained iron is able to absorb the blue flux sufficiently.
There are a few assumptions,
e.g. using a super-solar metallicity model or reducing the mass and/or radius
of the existing s9.0
model, which might result in the recombination wave receding faster (i.e. \emph{U}
magnitude to decline sharper). Magnitudes in \emph{V}, \emph{R}, and \emph{I} bands
tend to rise
after day~20, while observed SNe, particularly normal SNe~IIP or other low-luminosity 
SNe (as seen in \citealt{2014MNRAS.439.2873S}), have light curves 
declining in broad-bands \emph{B} and \emph{V}, or exhibit a plateau in \emph{R} and
\emph{I}.

There is another distinct feature seen in
Figure~\ref{figure:bands2005cs}: observed magnitudes have an 
significant flux excess compared to the synthetic light curves of the model s9.0. This
may indicate that the true progenitor is more extended than the model s9.0.
However, introducing a larger progenitor radius unavoidably leads to higher luminosity
during the plateau phase \citep{1993ApJ...414..712P}.

\subsection[SN~1999br]{SN~1999br}
\label{subsect:1999br}

\begin{figure}
\centering
\hspace{-5mm}\includegraphics[width=0.5\textwidth]{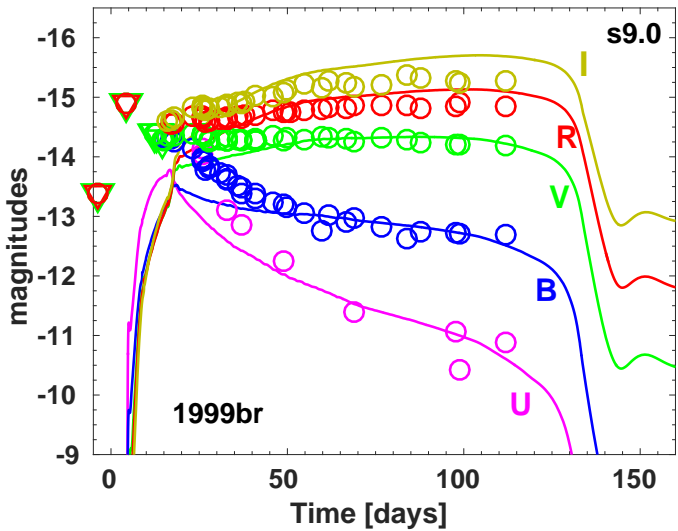}
\caption{Light curves for the model s9.0 and
observed SN~1999br in broad bands \citep{2004MNRAS.347...74P}.}
\label{figure:bands1999br}
\end{figure}

A number of low-luminosity SNe~IIP
have been observed \citep{2004MNRAS.347...74P,2014MNRAS.439.2873S}. We show
a comparison of
broad-band light curves of the model s9.0 to one of these SNe (SN~1999br) in Figure~\ref{figure:bands1999br}.
The distance to the parent galaxy of SN~1999br (NGC~4900) is subject to
uncertainty. We apply the up-to-date distance modulus from
\citet{2019MNRAS.485.1477K}. Accordingly, the magnitudes for all observed
broad-band magnitudes are
well matched by the model. However, there
is a flux excess at earlier time which suggests similarity between SN~1999br
and SN~2005cs. 
Another reason for the luminosity excess at
earlier time may lie in the possibility of interaction of the SN ejecta
with the circumstellar matter, expelled by the progenitor as wind
during earlier evolutionary
stages \citep{2017ApJ...838...28M,2018MNRAS.476.2840M,2020ApJ...895L..45G}.
However, adding a wind environment to the model leads to certain consequences for
observations. In particular, the SN can be expected to exhibit an 
X-ray excess which is not observed for every SN~IIP.

The low-luminosity SN~2008bk is considered to be similar to SN~1999br in
its photometric evolution \citep{2012AJ....143...19V}. 
According to \citet{2017MNRAS.466...34L}, SN~2008bk is best fitted by a
progenitor of initial mass 12~\Msun{} (model X in their study),
pre-explosion radius 502~\Rsun{} (consistent with
496~\Rsun{} derived by \citet{2004MNRAS.347...74P} and 500~\Rsun{} 
by \citet{2017MNRAS.464.3013P}), ejecta mass
of 8.29~\Msun{}, ejected $^{56}$Ni{} mass of 0.0086~\Msun{}, and explosion energy $2.5\times10^{\,50}$~erg.
However, \citet{2017MNRAS.466...34L} state that this energy is too high to match photospheric velocity
evolution \citep[see also][]{2018MNRAS.473.3863L}. In contrast, our model s9.0 fits
the observed velocity better due to
a significantly lower explosion energy of $0.68\times10^{\,50}$~erg and a lower
ejecta mass of 7.4~\Msun{} ($E/M=0.092$ in our study versus 0.3 in \citet{2017MNRAS.466...34L}).
Similarly, SN~1997D was explained by \citet{2000A&A...354..557C} as a
low-mass (6~\Msun{}) low energetic explosion ($1\times10^{\,50}$~erg).

\subsection[SN~2018zd and SN~2018hwm]{SN~2018zd and SN~2018hwm}
\label{subsect:2018}

It is worth mentioning that recently \citet{2020MNRAS.498...84Z}
and \citet{2020arXiv201102176H} presented evidence of SN~2018zd 
as an ECSN candidate.
However, the evolution of the broad-band magnitudes, specifically, the decline of the \emph{U} band
magnitude, is very similar to an average SN~IIP.
An analysis of spectra (FLASH-spectroscopy) 
and the time evolution of the photospheric velocity also indicate the
explosion of an average massive star in a surrounding wind \citep{2020MNRAS.498...84Z}. 
Hence, the most probable explanation is that SN~2018zd was a low- to intermediate-mass iron-CCSN. 
Note that the distance to the host galaxy (NGC\,2146) is estimated two times larger by 
\citet{2020MNRAS.498...84Z} than by \citet{2020arXiv201102176H}, which directly affects all
consequently derived values of the explosion energy, the mass of radioactive nickel {}$^{56}$Ni{}, 
radius of the progenitor and other parameters.
Specifically, \citet{2020MNRAS.498...84Z}
estimate 0.033~\Msun{} of {}$^{56}$Ni{} while \citet{2020arXiv201102176H}
calculate 0.0086~\Msun{} of {}$^{56}$Ni{}. Particularly, the latter
conclude that SN~2018zd was an ECSN based on the low mass of {}$^{56}$Ni{}.
Nevertheless, it is difficult to
discriminate different kinds of progenitors relying on the estimated {}$^{56}$Ni{}
mass, because the amount of {}$^{56}$Ni{}
produced during neutrino-driven explosions is mostly a function of the explosion
energy \citep{2016ApJ...821...38S,2020ApJ...890...51E}, 
and both ECSNe and low-mass iron-CCSNe can explode with similar energies. 

Moreover, \citet{2021MNRAS.501.1059R}
observed the low-luminosity SN~2018hwm and suggested two possibilities:
either it is an ECSN or a low-mass iron-CCSN. The authors do
not favour one possibility over the other. However, a low plateau luminosity does
not necessarily indicate an ECSN candidate, i.e. an explosion of a super-AGB star, 
because the luminosity depends on the radius,
and the radius of an ECSN progenitor might be very large. E.g. our extended ECSN
progenitor e8.8 exhibits a luminosity on the plateau of about
$9\times10^{\,41}$~erg\,s$^{\,-1}$, i.e. comparable to an average SN~IIP \citep{2002ApJ...566L..63H,2015ApJ...806..225P,2017ApJ...841..127M}, as we 
show in Figure~\ref{figure:bol}. On the other hand,
the very low photospheric velocity of about 1500~km\,s$^{\,-1}${} and very long plateau
of 150~days explicitly point to a low explosion energy, 0.055~foe, as derived by \citet{2021MNRAS.501.1059R}.
The very low mass of radioactive nickel {}$^{56}$Ni{}
of 0.002~\Msun{} also does not necessarily point the explosion as an
ECSN, because our low-mass iron-core explosions produce as little {}$^{56}$Ni{} as
0.0007~\Msun{} (model z9.6, see Table~\ref{table:values}). 
Therefore, SN~2018hwm is also most likely a low-mass iron-CCSN.

To conclude, the published observations cannot exclude
low-mass iron-CCSNe as an explanation of SN 2018zd and SN 2018hwm.

\subsection[B-V colour diagnostics]{\emph{B}\,--\emph{V} colour diagnostics}
\label{subsect:BmVobs}

\begin{figure}
\centering
\includegraphics[width=0.5\textwidth]{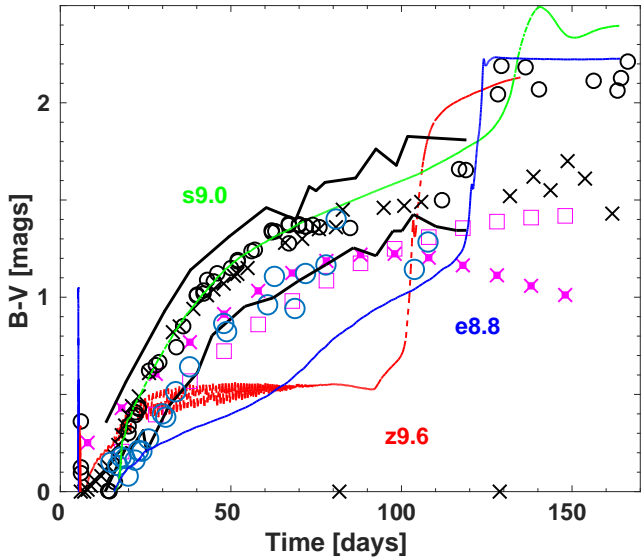}
\caption{The \emph{B}--\emph{V} colour for the models e8.8, z9.6 and s9.0 and
observed SN~1999em, SN~2013fs, ASASSN~2014gm, ASASSN~2014ha. We also superpose the lower and upper
limits for the \emph{B}--\emph{V} colour for normal SNe~IIP derived by
\citet{2014MNRAS.442..844F} represented by black lines.}
\label{figure:BmVsnIIP}
\end{figure}

In Figure~\ref{figure:BmVsnIIP}, we show the \emph{B}--\emph{V} colour for
our models e8.8, z9.6, and s9.0, in comparison to five normal SNe~IIP, namely
the typical SN~1999em, and also SN~2013fs, ASASSN~2014gm, ASASSN~2014ha, SN~2005cs
\citep{2009MNRAS.394.2266P,2014MNRAS.445..554F,2014MNRAS.442..844F,2016MNRAS.459.3939V}.  
As mentioned by \citet{2019MNRAS.483.1211K}, the \emph{B}--\emph{V} colour
could serve as an indicator of the $^{56}$Ni{} mixing in the SN ejecta. Indeed, the models
e8.8 and z9.6 have a strictly stratified chemical structure, leading to a
sharp reddening at the end of the plateau, while the model s9.0
has extensive mixing of radioactive nickel in the ejecta. Extensive mixing of radioactive nickel
leads to colour reddening more
gradually at the end of the plateau. Still, none of the models in our study
matches the behaviour of the \emph{B}--\emph{V} colour for the typical SNe~IIP
in our sample. It is possible that the model s9.0 might explain the colour of
SN~2005cs while having a smaller mass of $^{56}$Ni{}. With 0.005~\Msun{} of $^{56}$Ni{}
the colour reddens to much after the transition to the tail, meaning the
absorption, i.e. line opacity, is very strong in the inner ejecta.

Contrary to the conclusions of \citet{2009MNRAS.398.1041B}, the light curve of SN~2008S 
is dissimilar  to the ECSN model e8.8. 
The distinguishing property of an ECSN, i.e. a super-AGB progenitor explosion, is
not a low-luminosity plateau, but a distinct moderate luminosity plateau
accompanied by a sharp transition to a low-luminosity radioactive tail.
\citet{2013ApJ...771L..12T}
also speculate that SN~2008S is most likely not an ECSN explosion. However,
they suggest that a more compact ECSN progenitor might be a good candidate for
SN~2008S, especially taking the uncertainty of the wind mass-loss and
pulsationally-driven mass-loss into account.

The sub-class of very faint, faint, or/and low-luminosity SNe~IIP
\citep{2004MNRAS.347...74P,2007Natur.449E...1P,2014MNRAS.439.2873S}
are most likely explosions of low-mass progenitors, similar to our model
z9.6 but at solar metallicity and a lower energy, e.g. a few
$10^{\,49}$~erg, and relatively low masses of ejected $^{56}$Ni{}. The very low
energies of these explosions are favoured by the low photospheric velocity
during the plateau phase.


\section[Conclusions]{Conclusions}
\label{sect:conclusions}

In this study, we present the evolution of the SN ejecta of
three progenitors: the ECSN progenitor e8.8, the low-mass zero-metallicity
z9.6, and the low-mass solar-metallicity s9.0, with initial ZAMS masses
of 8.8~\Msun{}, 9.6~\Msun{}, and 9~\Msun{}, respectively. The models were
exploded self-consistently with \verb|PROMETHEUS| by
\citet{2020MNRAS.496.2039S}. We calculated the hydrodynamical evolution and
radiative transfer simulations for these three models with \verb|STELLA|.
The resulting light curves differ from explosions of massive stars. We used
15~\Msun{} model L15 \citep{2000ApJS..129..625L} as a reference CCSN model for comparison.
Among the reasons for the differences are:
\begin{itemize}
\item low explosion energy;
\item  low mass of ejected radioactive nickel $^{56}$Ni{};
\item absence of the distinct massive ($>1$~\Msun{}) helium shell and oxygen
layer.
\end{itemize}

The distinct properties of the SNe arising from our default models
can be summarised as:

\underline{\textbf{The model e8.8:}}
\begin{itemize}
\item during the first 50~days: plateau in \emph{U} band and rising
magnitudes in \emph{B}, \emph{V}, \emph{R}, and \emph{I} bands.
\item after day~50: plateau in \emph{R} and \emph{I} bands.
\item the transition to the tail is steep and pronounced, more than two orders in
bolometric luminosity, or 6~mags in \emph{V} band.
\item colour temperature remains above 6000~K until the middle of the plateau.
\end{itemize}

\underline{\textbf{The model z9.6:}}
\begin{itemize}
\item relatively low luminosity on the plateau, $\log
L_\mathrm{bol}\sim10^{\,41}$~erg\,s$^{\,-1}$, or --14.5~mags in \emph{V} band.
\item pronounced and steep transition to the tail, about 1.5 orders in
bolometric luminosity, or 4.6~mags in \emph{V} band.
\item magnitudes in all broad bands increase starting at early times. The 
colours remain largely constant during the plateau phase.
\end{itemize}

\underline{\textbf{The model s9.0:}}
\begin{itemize}
\item relatively low luminosity during the plateau phase, $\log
L_\mathrm{bol}\sim10^{\,41}$~erg\,s$^{\,-1}$, or --14~mag in \emph{V} band.
\item shallow transition to the tail due to extended macroscopic mixing
of radioactive material.
\item the transition is less pronounced, about 3.5~mags in \emph{V} band.
\end{itemize}

The model s9.0 is the best candidate among our models for the observed low-luminosity
SNe~IIP according to its broad-band light curves and colour evolution.

For all our models, photospheric velocity is relatively low, 1000\,--\,2000~km\,s$^{\,-1}$
during the plateau. This feature is a convenient indicator of low-mass
explosions.

We do not find a good candidate among our sample of observed SNe
resembeling the 
observables of the ECSN model e8.8. According to our models, low-luminosity SNe~IIP
like SN~1999br and SN~2005cs, 
can be explained by the explosion of low-mass CCSNe. However, the observed
SNe show a clear flux excess at the earlier phase (before day~20). 
This requires either more extended progenitors or the presense of circumstellar
matter (e.g. pre-SN wind), and interaction of the SN ejecta with the
circumstellar environment.

Further we carried out the following studies:
\begin{enumerate}
\item Variations of the $^{56}$Ni{} mass produced in model e8.8. The more $^{56}$Ni{}
is ejected, the longer is
the plateau and the higher the luminosity on
the tail. Slightly bluer colours (0.2~mags in \emph{B}--\emph{V}).
\item Variations of the explosion energy for the model e8.8. We find that the light curves obey the
standard relations from \citet{1993ApJ...414..712P}, i.e. the higher the energy
the shorter and brighter the plateau. Slightly bluer colours
(0.2~mags in \emph{B}--\emph{V}).
\item Variations of the hydrogen-to-helium ratio in the model e8.8. Here we find that the higher helium fraction
in the hydrogen-rich envelope, the shorter and brighter the plateau. Slightly bluer colours
(0.2~mags in \emph{B}--\emph{V}).
\item A metallicity study for the models e8.8 and z9.6. We find that 
the higher the metallicity, i.e. the higher the iron abundance in
the hydrogen-rich envelope, the redder the colours; the \emph{U} band magnitude
is good indicator for measuring the metallicity of the SN progenitor; 
the \emph{B}--\emph{V} colour
changes significantly: 1~mag between zero metallicity case and
SMC/solar metallicity.
\item A radius-dependence study for the ECSN model e8.8. Assuming the ECSN
explodes in a binary system, the progenitor may lose hydrogen-rich envelope
via close binary interaction. We found out that the light curves for the
truncated models become shorter, namely, the sub-model with the radius
400~\Rsun{} has the sharply declining 30~day light curve with a
low-luminosity maximum phase. Therefore, ECSNe in binaries are mostly undetectable.
\end{enumerate}

Spectral synthesis simulations for our models similar to \citet{2018MNRAS.475..277J} 
will be useful, as synthetic spectra
are more sensitive to the SN ejecta structure than broad-band
light curves. 

\section*{Acknowledgments}
AK is supported by the Alexander von Humboldt Foundation. 
PB is sponsored by  grant  RFBR 21-52-12032 in his work on the \verb|STELLA| code development.
HTJ acknowledges support by the Deutsche Forschungsgemeinschaft (DFG, German
Research Foundation) through Sonderforschungsbereich (Collaborative Research
Center) SFB-1258 ``Neutrinos and Dark Matter in Astro- and Particle Physics (NDM)'' and under Germany's
Excellence Strategy through Cluster of Excellence ORIGINS (EXC-2094)-390783311, and by the European Research
Council through Grant ERC-AdG No.~341157-COCO2CASA. The authors thank
Andrea Pastorello for providing the observed data in a suitable format and corresponding discussions.
AK would like to thank Patrick Neunteufel for useful suggestions.

\addcontentsline{toc}{section}{Acknowledgments}

\section*{Data availability}

The data computed and analysed for the current study are available via link
\url{https://wwwmpa.mpa-garching.mpg.de/ccsnarchive/data/Kozyreva2018/}.
Results of the core-collapse explosion simulations are available for 
download upon request on the following website:
\url{https://wwwmpa.mpa-garching.mpg.de/ccsnarchive/archive.html} .

\bibliographystyle{mnras}





\input{ecsn2.bbl}




%


\bsp	
\label{lastpage}
\end{document}